\null

\input epsf

\magnification 1200


\newskip\ttglue

\font\eightrm=cmr8
\font\eighti=cmmi8
\font\eightsy=cmsy8
\font\eightbf=cmbx8
\font\eighttt=cmtt8
\font\eightsl=cmsl8
\font\eightit=cmti8
\font\sixrm=cmr6
\font\sixbf=cmbx6
\font\sixi=cmmi6
\font\sixsy=cmsy6

\def \eightpoint{\def\rm{\fam0\eightrm}
\textfont0=\eightrm \scriptfont0=\sixrm \scriptscriptfont0=\fiverm
\textfont1=\eighti \scriptfont1=\sixi \scriptscriptfont1=\fivei
\textfont2=\eightsy \scriptfont2=\sixsy \scriptscriptfont2=\fivesy
\textfont3=\tenex \scriptfont3=\tenex \scriptscriptfont3=\tenex
\textfont\itfam=\eightit \def\it{\fam\itfam\eightit}%
\textfont\slfam=\eightsl \def\sl{\fam\slfam\eightsl}%
\textfont\ttfam=\eighttt \def\tt{\fam\ttfam\eighttt}%
\textfont\bffam=\eightbf \scriptfont\bffam=\sixbf
\scriptscriptfont\bffam=\fivebf \def\bf{\fam\bffam\eightbf}
\ttglue=.5em plus.25em minus.15em
\setbox\strutbox=\hbox{\vrule height7pt depth2pt width0pt}%
\normalbaselineskip=9pt
\let\sc=\sixrm \let\big=\eightbig \normalbaselines\rm }
\def\a{\alpha}
\def\b{\beta}
\def\c{\gamma}
\def\d{\delta}
\def\e{\epsilon}

\def\l{\lambda}
\def\m{\mu}
\def\n{\nu}
\def\o{\theta}

\def\s{\sigma}
\def\t{\tau}

\def\C{\Gamma}
\def\D{\Delta}
\def\L{\Lambda}

\def\pl{\partial}

\def\DD{{\cal D}}

\def\GV{{\rm GeV}}
\def\MV{{\rm MeV}}

\def\as{{\alpha_s}}
\def\asp{\Bigl({\alpha_s\over\pi}\Bigr)} \def\asb{{\bar\alpha}_s}
\def\aspb{\Bigl({{\bar\alpha}_s\over\pi}\Bigr)}   
\def\aQ{\bigl(\a_s(Q^2)\bigr)}
\def\msb{\overline{MS}}
\def\bm{{\bar m}_s}
\def\th{\vartheta}


{\nopagenumbers

\line{\hfil CERN--TH/98--385 }
\line{\hfil PM/98-37 }
\line{\hfil SWAT 98/208 }
\vskip0.5cm
\centerline{\bf TOPOLOGICAL CHARGE SCREENING AND THE `PROTON SPIN'}
\vskip0.4cm
\centerline{\bf BEYOND THE CHIRAL LIMIT}
\vskip1.2cm
\centerline{\bf S.~Narison${}^*$, G.M. Shore${}^{**}$ and G. Veneziano${}^{***}$}
\vskip0.7cm
\centerline{\it ${}^*$ Laboratoire de Physique Math\'ematique et Th\'eorique} 
\centerline{\it Universit\'e de Montpellier II}
\centerline{\it Place Eug\'ene Bataillon} 
\centerline{\it F34095 Montpellier ~Cedex 05, France} 
\vskip0.5cm
\centerline{\it ${}^{**}$ Department of Physics} 
\centerline{\it University of Wales Swansea} 
\centerline{\it Singleton Park}
\centerline{\it Swansea, SA2 8PP, U.K. } 
\vskip0.5cm
\centerline{\it ${}^{***}$ Theory Division} 
\centerline{\it CERN}
\centerline{\it CH 1211 Geneva 23, Switzerland} 
\vskip0.7cm
\noindent{\bf Abstract}
\vskip0.3cm
The theory of the `proton spin' effect proposed in our earlier papers
is extended to include the chiral $SU(3)$ symmetry breaking and
flavour mixing induced by non-vanishing quark masses in QCD. 
The theoretical basis is the derivation of exact, unified 
Goldberger-Treiman (GT) relations valid beyond the chiral limit.
The observed suppression in the flavour singlet axial charge $a^0(Q^2)$
is explained by an anomalously small value for the slope of the
singlet current correlation function $\langle 0|T~\pl^\m J_{\m 5}^0 ~
\pl^\n J_{\n 5}^0 |0\rangle$, a consequence of the screening
of topological charge in the QCD vacuum.
Numerical predictions are obtained by evaluating the current correlation
functions using QCD spectral sum rules. The results,
$a^0(Q^2) = 0.31 \pm 0.02$ and $\int dx ~g_1^p(x,Q^2) = 0.141 \pm 0.005$
(at $Q^2=10 ~\GV^2$), are in good agreement with current experimental
data on the polarised proton structure function $g_1^p$.

\vskip0.7cm

\line{CERN--TH/98--385 \hfil}
\line{PM/98-37	\hfil}
\line{SWAT 98/208 \hfil}
\vskip0.1cm
\line{December, 1998 \hfil}

\vfill\eject }

\pageno = 1

\noindent {\bf 1. Introduction}
\vskip0.5cm
The `proton spin' problem, i.e.~the question of why the first moment of the
flavour singlet component of the polarised proton structure function $g_1^p$ 
is anomalously suppressed, has inspired an impressive research effort, both 
theoretical and experimental, for over a decade. (For recent reviews, see 
e.g.~refs.[1,2]. As is well-known, the first 
moment of $g_1^p$ can be expressed in terms of the axial charges of the 
proton as follows:
$$
\int_0^1 dx~ g_1^p(x;Q^2) ~=~
{1\over12} C_1^{\rm NS}\aQ~ \Bigl( a^3 + {1\over3} a^8 \Bigr) +
{1\over9} C_1^{\rm S}\aQ~ a^0(Q^2)
\eqno(1.1)
$$
Here, $C_1^{\rm NS}$, $C_1^S$ are the appropriate Wilson coefficients arising from
the OPE for two electromagnetic currents, while the axial charges are defined
from the forward matrix elements
$$
\langle p,s|J_{\m 5}^3|p,s\rangle = {1\over2} a^3 s_\m  ~~~~~~~
\langle p,s|J_{\m 5}^8|p,s\rangle = {1\over{2\sqrt3}} a^8 s_\m  ~~~~~~~
\langle p,s|J_{\m 5}^0|p,s\rangle = a^0(Q^2) s_\m
\eqno(1.2)
$$
where $J_{\m 5}^a$ (often denoted $A_{\m}^a$) are the axial currents 
and $s_\m$ is the proton polarisation vector.
$a^3$ and $a^8$ are known in terms of the $F$ and $D$ coefficients from
beta and hyperon decay, so that an experimental determination of the
first moment of $g_1^p$ in polarised deep inelastic scattering (DIS) allows a 
determination of the singlet axial charge $a^0(Q^2)$.
The `proton spin' problem is that it is found experimentally that
$a^0(Q^2)$ is strongly suppressed relative to $a^8$, which would be its 
expected value if the OZI (Zweig) rule were exact in this channel.

\vskip0.2cm
DIS is normally described theoretically using the QCD parton model.
In this model, the axial charges are represented (in the AB renormalisation
scheme) in terms of moments of parton distributions as follows[3,4]:
$$
a^3 = \D u - \D d ~~~~~~~~
a^8 = \D u + \D d - 2\D s ~~~~~~~~
a^0(Q^2) = \D u + \D d + \D s - n_f {\a_s\over2\pi} \D g(Q^2)
\eqno(1.3)
$$
In the parton model, the `proton spin' problem takes the following 
form. In the naive, or valence quark, parton model we would expect the
strange quark and gluon distributions to vanish, i.e.~$\D s = 0$,
$\D g(Q^2) = 0$. In that case, $a^0 = a^8$, the OZI prediction.
Inserted into eq.(1.1), this gives the Ellis-Jaffe sum rule[5]. However,
the observed suppression $a^0(Q^2) < a^8$ can be accommodated in the 
full QCD parton model by invoking either or both a non-zero polarised 
strange quark distribution $\D s \neq 0$ or a non-zero polarised gluon
distribution $\D g(Q^2) \neq 0$. An interesting conjecture (in line with
the insights of our alternative approach) is that the suppression 
is primarily due to the gluon distribution[3], although a quantitative prediction
would still only follow if $\D g(Q^2)$ can be independently measured, either
through a precise analysis of the $Q^2$ dependence of $g_1^p$ [4] or directly
through other less inclusive high energy processes such as open charm production.
Notice, however, that even in the QCD parton model picture, it is {\it not}
possible to identify $a^0(Q^2)$ with spin[6]. This identification only holds
for free quarks, in which case the $Q^2$ scale dependence (which is related
to the $U_A(1)$ axial anomaly) disappears from $a^0$. As has been emphasised
many times[7-9,1], the so-called `proton spin' problem is not a problem of 
{\it spin} -- rather, it is a question of understanding the dynamical 
origin of the OZI violation $a^0(Q^2) < a^8$. 

\vskip0.2cm
In a series of papers[7,10-12], we have a proposed an alternative, complementary, 
approach to the `proton spin' problem which provides both a new physical insight
into the origin of the suppression in $a^0(Q^2)$ and a quantitative 
prediction, which is in good agreement with the current experimental data.
In our approach (reviewed in refs.[8,1]), the flavour singlet axial charge 
$a^0(Q^2)$ is decomposed into
the product of an RG-invariant 1PI vertex and a non-perturbative, but target 
independent, QCD correlation function which we subsequently identify as the 
first moment of the QCD topological susceptibility $\chi'(0)$, i.e.
$$
a^0(Q^2) ~~=~~ {1\over 2m_N} 6 ~\sqrt{\chi'(0)} ~\hat\Gamma_{\eta^0 NN}
\eqno(1.4)
$$
where $\chi'(0) = {d\over dk^2}\chi(k^2)|_{k=0}$ and
$$
\chi(k^2) = i\int d^4x~e^{ikx}\langle 0|T~Q(x)~Q(0)|0\rangle
\eqno(1.5)
$$
with $Q = {\a_s\over8\pi} {\rm tr} G_{\m\n} \tilde G^{\m\n}$ the gluon 
topological charge density. In the vertex, $\eta^0$ denotes the `OZI 
Goldstone boson', i.e.~the (unphysical) state which would become the Goldstone 
boson for spontaneously broken $U_A(1)$ in the absence of the anomaly 
(OZI limit). 
We now make the key assumption that the RG-invariant vertex obeys the
OZI rule, viz. $\hat\Gamma_{\eta^0 NN}
= \sqrt2 ~\hat\Gamma_{\eta^8 NN}$. 
Then, comparing with the GT relation for
$a^8$ in the chiral limit, we obtain our prediction:
$$
{a^0(Q^2)\over a^8} ~=~ {\sqrt6 \over f_\pi}~
\sqrt{\chi'(0)}\Big|_{Q^2}
\eqno(1.6)
$$
The suppression in $a^0(Q^2)$ simply reflects
an anomalously small value of $\chi'(0)$, which we confirmed by an 
explicit calculation using QCD spectral sum rules[12]. In this picture,
the suppression is therefore a target-independent effect, i.e.~a generic
property of the QCD vacuum rather than a specific property of the proton.
Since $a^0(Q^2)$ can be expressed, via the axial $U_A(1)$ anomaly, as
the proton matrix element of the topological charge $Q$, its suppression
can be understood as a screening of topological charge in the QCD vacuum.
We conclude that the fundamental dynamics underlying the violation
of the Ellis-Jaffe sum rule is not to do with quark spin, but instead is
a manifestation of topological charge screening by the QCD vacuum.

This `target-independence' property may in principle be tested directly
in experiment, by studying semi-inclusive DIS in which a hadron carrying a
large fraction of the target energy is observed in the target fragmentation
region. Such experiments should be possible at, e.g.~polarised HERA.
The details of our proposal are described in refs.[13-15].

\vskip0.2cm
Although our previous analysis was restricted to the chiral limit of QCD,
it was clear from the derivation that the final result for $a^0(Q^2)$
should only have a weak dependence on the quark masses. 
Indeed, the cancellation of the explicit quark mass dependence in $a^0(Q^2)$
was already pointed out in ref.[7]. However, the 
introduction of quark masses does involve a significant complication in 
the analysis. The purpose of this paper is therefore to present the 
generalisation of our theory of the `proton spin' to QCD beyond the chiral limit.

Underlying our approach to polarised DIS -- the `CPV' method reviewed in
some detail in ref.[1] -- is the $U_A(1)$ Goldberger-Treiman (GT) relation.
The initial insight that the `proton spin' problem was essentially one of
OZI violation and could be understood in terms of an extension of the
GT relation to the flavour singlet channel was made
by one of us in ref.[7]. The $U_A(1)$ GT relation was 
subsequently put on a firm field-theoretical basis in refs.[10,11],
where the connection with the slope of the topological susceptibility was 
first established.

The introduction of quark masses complicates the $U_A(1)$ GT analysis
in two ways. First, the pure gluon topological susceptibility has
to be combined with correlation functions of pseudoscalar operators of the 
form $\phi_5 = \sum \bar q \c_5 q$ arising from the extra explicit chiral symmetry
breaking terms in the divergence of the axial current. Second, since 
the quark masses break flavour $SU(3)$, the $U_A(1)$ GT relation is mixed with 
those for the flavour non-singlet axial charges. In this paper, we
present the unified GT relations for the flavour singlet and non-singlet
axial charges, expressing them in terms of 1PI vertices and the first
moments of the correlation functions of the divergences of the
axial currents as follows:
$$
G_A^a ~=~ {1\over 2m_N}~F_{ab}~\hat\Gamma_{\eta^b NN}
\eqno(1.7)
$$
where $a=0,3,8$ is the $SU(3)$ flavour index, the axial charges are
normalised as
$$
G_A^3 = {1\over2} a^3 ~~~~~~~~
G_A^8 = {1\over 2\sqrt3} a^8 ~~~~~~~~
G_A^0(Q^2) = a^0(Q^2)
\eqno(1.8)
$$
and
$$
(FF^T)_{ab} ~~=~~ \lim_{k=0} {d\over dk^2}~i\int d^4 x~
\langle 0|T~\pl^\m J_{\m 5}^a(x)~\pl^\n J_{\n 5}^b(0) |0\rangle 
\eqno(1.9)
$$
It is important to realise that, like the singlet
$U_A(1)$ GT relation, these are {\it exact} relations in QCD field theory.
They go beyond the pole-dominance approximation used in the conventional
GT relations, replacing the decay constants by current correlation functions
and the $\pi NN$ or $\eta NN$ couplings by appropriate 1PI vertex functions.

In the second part of the paper, we make this analysis quantitative by 
calculating the relevant current correlation functions using QCD spectral
sum rules. This confirms the stability of our earlier results[12] to the 
introduction of quark masses. A very careful analysis is presented, with
particular attention given to the inclusion of higher order corrections
in $\a_s$ or $m_s^2 \t$ and to the stability of the sum rules with respect to the
Borel parameter $\t$. We find that the size of the explicit chiral and $SU(3)$
breaking effects is in line with our theoretical expectations and confirms
the validity of the Laplace sum rule method for calculating correlation
functions, such as the topological susceptibility, in the flavour singlet
channel.

\vskip0.2cm
The paper is organised as follows. In section 2, we quote the (anomalous)
chiral Ward identities which we need to derive the unified 
Goldberger-Treiman relations. The derivation itself is presented
in section 3. In section 4, we relate these new GT relations to the
`proton spin' sum rule and describe our expectations for the results
of a spectral sum rule calculation of the relevant current correlation
functions. This is elaborated on in section 5, where the pattern
of explicit mass cancellations is exhibited in the framework of
an effective lagrangian.

In section 6, we evaluate the current correlation functions (1.9) using 
the QCD spectral sum rule technique. The presence of quark masses
requires a substantially more complicated analysis than we had previously
performed in the chiral limit to evaluate the slope of the topological
susceptibility. However, the results confirm our earlier conclusions[12].
In section 7, these numerical results are used in the
GT relations and our prediction for the `proton spin' suppression is
obtained. Our conclusions from this work are presented in section 8.

Appendix A deals with some technical aspects of the derivation of the
unified GT formulae. Their relationship with conventional
current algebra and PCAC methods is discussed briefly in Appendix B. 
The renormalisation group properties of the Green functions and vertices
involved in the GT relations are derived in Appendix C.
Finally, since both our formal methods
and spectral sum rule analysis have been repeatedly criticised in the
literature by Ioffe (see e.g.~refs.[16,17] and references therein),
we explain in Appendix D why these criticisms are not correct
and show in detail the errors in refs.[16-20] which lead Ioffe to 
his false conclusions.

\vskip1cm

\noindent{\bf 2. Chiral Ward Identities} 
\vskip0.5cm
The derivation of the generalised Goldberger-Treiman relations is based on
the chiral Ward identities satisfied by the composite operator propagators
and vertex functions. In this section, we derive these in the form which
will be most convenient for the applications which follow.

For the propagators, the starting point is the Ward identity for the
generating functional $W[V_{\m5}^a, V_\m^a, \o, S_5^a, S^a]$ of Green
functions which are `1PI' with respect to the designated fields (composite
operators).
Here, $V_{\m5}^a, V_\m^a, \o, S_5^a, S^a$ are the sources for the composite
operators $J_{\m5}^a, J_\m^a, Q, \phi_5^a, \phi^a$ respectively, where
$$\eqalignno{
J_{\m5}^a &= \bar q \c_\m \c_5 T^a q
~~~~~~~~~~~~~~~~~
J_\m^a = \bar q \c_\m T^a q
~~~~~~~~~~~~~~~~~
Q = {\a_s\over8\pi} {\rm tr} G_{\m\n} \tilde G^{\m\n} \cr \phi_5^a &= \bar
q \c_5 T^a q
~~~~~~~~~~~~~~~~~~~~~
\phi^a = \bar q T^a q
&(2.1) \cr }
$$
In this notation, $T^i = {1\over2}\l^i$ are flavour $SU(n_f)$ generators,
and we include the singlet $U_A(1)$ generator $T^0 = {\bf 1}$ and let the
index $a = 0, i$.
We will only need to consider fields where $i$ corresponds to a generator
in the Cartan sub-algebra, so that $a = 0, 3, 8$ for $n_f = 3$ quark
flavours. We define $d$-symbols by $\{T^a,T^b\} = d_{abc} T^c$. Since this
includes the flavour singlet $U(1)$ generator, they are only symmetric on
the first two indices. For $n_f = 3$, the explicit values are $d_{000} =
d_{033} = d_{088} = 2, d_{330} = d_{880} = 1/3, d_{338} = d_{383} =
-d_{888} = 1/\sqrt3$. (For further notation and description of the
formalism used here, see ref.[11].)

The chiral Ward identities are written in the functional formalism as
follows: $$
\pl_\m {\d W\over\d V_{\m5}^a} - 2n_f \d_{a0} {\d W\over\d\o} - d_{adc} m^d
{\d W\over \d S_5^c}
+ d_{adc} S^d {\d W \over \d S_5^c}
- d_{adc} S_5^d {\d W\over\d S^c}
= 0
\eqno(2.2)
$$
where we have displayed both the anomalous breaking term for the $U_A(1)$
current and the soft breaking induced by the quark masses. The quark mass
matrix is written as $m^a T^a$, so that for $n_f=3$, $$
\left(\matrix{m_u &0 &0 \cr
0 &m_d &0 \cr
0 &0 &m_s \cr}\right)
= m^0 {\bf 1} + m^3 T^3 + m^8 T^8
\eqno(2.3)
$$
Notice that there are no variation terms for the currents themselves as we
have restricted to fields where $a$ is a Cartan sub-algebra index. It is
convenient to use a notation where a functional derivative is denoted
simply by a subscript. So, also transforming to momentum space, we can
rewrite eq.(2.2) compactly as
$$
ik_\m W_{V_{\m5}^a} - 2n_f \d_{a0} W_{\o} - d_{adc} m^d W_{S_5^c} + d_{adc}
S^d W_{S_5^c} - d_{adc} S_5^d W_{S^c}
= 0
\eqno(2.4)
$$

The Ward identities for composite operator Green functions are derived by
taking functional derivatives of this basic identity. We will need the
following identities for 2-point functions:
$$\eqalignno{
&ik_\m W_{V_{\m5}^a V_{\n5}^b} - 2n_f \d_{a0} W_{\o V_{\n5}^b} - M_{ac}
W_{S_5^c V_{\n5}^b} = 0 \cr
&ik_\m W_{V_{\m5}^a \o} - 2n_f \d_{a0} W_{\o \o} - M_{ac} W_{S_5^c \o} = 0 \cr
&ik_\m W_{V_{\m5}^a S_5^b} - 2n_f \d_{a0} W_{\o S_5^b} - M_{ac} W_{S_5^c
S_5^b} - \Phi_{ab} = 0 &(2.5) \cr }
$$
where we have introduced the still more compact notation $$
M_{ab} = d_{acb} m^c
~~~~~~~~~~~~~~~~~~
\Phi_{ab} = d_{abc} \langle \phi^c\rangle \eqno(2.6)
$$
$\langle \phi^c\rangle$ is the VEV $\langle \bar q T^c q\rangle$, so that
for $n_f=3$ the condensates may be written as $$
\left(\matrix{ \langle \bar u u\rangle &0 &0 \cr 0 &\langle \bar d d\rangle
&0 \cr
0 &0 &\langle \bar s s\rangle \cr}\right) = {1\over3} \langle\phi^0\rangle
{\bf 1} + 2\langle\phi^3\rangle T^3 + 2 \langle\phi^8\rangle T^8 
\eqno(2.7)
$$
Notice that with these definitions,
$$
{1\over8} {\rm det} M = m_u m_d m_s
~~~~~~~~~~~~~
{1\over6} {\rm det} \Phi =
\langle \bar u u \rangle ~\langle \bar d d \rangle ~\langle \bar s s
\rangle \eqno(2.8)
$$
with the obvious generalisation for arbitrary $n_f$.

Combining the individual equations in (2.5), we find the important identity: $$
k_\m k_\n W_{V_{\m5}^a V_{\n5}^b} - M_{ac} \Phi_{cb} = W_{S_D^a S_D^b}
\eqno(2.9)
$$
where $S_D^a$ is the source for the current divergence operator $D^a = 2n_f
\d_{a0} Q + d_{acd} m^c \phi_5^d$. In conventional notation,
$$\eqalignno{
W_{S_D^a S_D^b} &= i \int dx~e^{ikx}~\langle 0| T~D^a(x)~D^b(0)|0\rangle
\cr &= i \int dx~e^{ikx}~\langle 0| T~\pl^\m J_{\m5}^a(x)~\pl^\n
J_{\n5}^b(0)|0\rangle &(2.10) \cr }
$$

The zero-momentum Ward identities play a special role. These follow
immediately from eqs.(2.5) under the assumption that there are no massless
particles (in particular, no exact Goldstone bosons) contributing $1/k^2$
poles in the 2-point functions. With this assumption, we find simply
$$\eqalignno{
&2n_f \d_{a0} W_{\o\o} + M_{ac} W_{S_5^c \o} = 0 \cr &2n_f \d_{a0} W_{\o
S_5^b} + M_{ac} W_{S_5^c S_5^b} + \Phi_{ab} = 0 &(2.11) \cr }
$$

The generating functional for proper vertices, $\C$, is defined from $W$ by
a partial Legendre transform (Zumino transform[21]), in which the
transform is made only on the fields $Q, \phi_5^a, \phi^a$ and not on the
currents. A number of important results on these transforms are collected
in Appendix A. The resulting proper vertices are `1PI' wrt the propagators
for these composite operators only. As explained fully in ref.[11], by
separating off the particle poles in the propagators, this is the
definition which gives the closest identification of these field-theoretic
vertices with physical low-energy couplings such as $g_{\pi NN}$ etc.

We therefore define the generating functional $\C[V_{\m5}^a, V_\m^a, Q,
\phi_5^a, \phi^a]$ as:
$$
\C[V_{\m5}^a, V_\m^a, Q, \phi_5^a, \phi^a] = W[V_{\m5}^a, V_\m^a, \o,
S_5^a, S^a] -
\int dx~\Bigl(\o Q + S_5^a \phi_5^a + S^a \phi^a \Bigr) \eqno(2.12)
$$

The chiral Ward identities corresponding to eq.(2.4) are therefore: $$
ik_\m \C_{V_{\m5}^a} - 2n_f \d_{a0} Q - d_{acd} m^c \phi_5^d + d_{acd}
\phi^d \C_{\phi_5^c} - d_{acd} \phi_5^d \C_{\phi^c} = 0 \eqno(2.13)
$$
The Ward identities for the 2-point vertices will also be important in the
derivation of the GT relations. These follow directly from eq.(2.13):
$$\eqalignno{
&ik_\m \C_{V_{\m5}^a V_{\n5}^b} + \Phi_{ac} \C_{\phi_5^c V_{\n5}^b} = 0 \cr
&ik_\m \C_{V_{\m5}^a Q} - 2n_f \d_{a0} + \Phi_{ac} \C_{\phi_5^c Q} = 0 \cr
&ik_\m \C_{V_{\m5}^a \phi_5^b} + \Phi_{ac} \C_{\phi_5^c \phi_5^b} - M_{ab}
= 0 &(2.14) \cr }
$$
It is then straightforward to derive the following important identity,
analogous to eq.(2.9):
$$
k_\m k_\n \C_{V_{\m5}^a V_{\n5}^b} + M_{ac} \Phi_{cb} = \Phi_{ac}
\C_{\phi_5^c \phi_5^d} \Phi_{db} \eqno(2.15)
$$

\vfill\eject

\noindent {\bf 3. The Goldberger-Treiman Relations} \vskip0.5cm
We now present a unified derivation of the Goldberger-Treiman relations for
the flavour singlet and non-singlet axial charges.\footnote{$\eightpoint
{}^{(1)}$}{\eightpoint 
\noindent Similar generalised GT relations have been presented some time
ago, with the same motivation of explaining the `proton spin' problem,
by Efremov, Soffer and T\"ornqvist[22] using the more
conventional PCAC language.} The derivation follows
the principles of refs.[10,11], extended to include non-zero quark masses
and flavour mixing. This allows us to present all the GT relations in terms
of a single, unified formula involving the 2-point correlation functions of
the divergences of the flavour singlet and non-singlet currents.

The axial charges $G_A^a$ are defined as the form factors in the forward
nucleon matrix elements of the axial currents, viz. $$
\langle p,s|J_{\m5}^a|p,s\rangle = G_A^a s_\m \eqno(3.1)
$$
where $p_\m$ and $s_\m = \bar u(p,s)\c_\m \c_5 u(p,s)$ are respectively
the momentum and polarisation vector of the nucleon.

To express this matrix element in our composite propagator-vertex
formalism, we introduce an interpolating field $N$ and source $S_N$ for the
nucleon as in refs.[10,11]. (Notice that this is purely a formal device --
there is no dynamics implicit in this manoeuvre.) The matrix element is
then just the 3-point function $W_{V_{\m5}^a S_N S_N}$ with the external
propagators amputated. From eq.(A.8), we can re-express this in terms of
the vertex functional $\C$ as follows:
$$\eqalignno{
\langle p,s|J_{\m5}^a|p,s\rangle ~&=~ \bar u(p,s) \biggl[
W_{S_N S_N}^{-1}~ W_{V_{\m5}^a S_N S_N}~ W_{S_N S_N}^{-1}\biggr] 
u(p,s) \cr &{}\cr
&= ~\bar u(p,s)\biggl[\C_{V_{\m5}^a NN} ~+~ W_{V_{\m5}^a \o}~ 
\C_{QNN}~ +~ W_{V_{\m5}^a S_5^b}~ \C_{\phi_5^b NN}\biggr] u(p,s) &(3.2) \cr }
$$
Since the propagators on the rhs vanish at zero momentum (this requires the
absence of any $1/k^2$ poles, which is assured by the $U_A(1)$ anomaly and
quark masses), we find simply
$$
2m_N G_A^a~\bar u \c_5 u = \bar u \biggl[k_\m \C_{V_{\m5}^a NN} \big|_{k=0}
\biggr] u \eqno(3.3)
$$
where $m_N$ is the nucleon mass. The GT relations then follow immediately
from the Ward identity (2.13) for $\C$. Differentiating wrt the nucleon
fields, we find
$$
2m_N G_A^a =  \Phi_{ab} \hat\C_{\phi_5^b NN} \big|_{k=0} \eqno(3.4)
$$
where for convenience we define $i\bar u ~\C_{\phi_5^a NN}~ u = 
\hat\C_{\phi_5^a NN} ~\bar u \c_5 u$, etc.

A non-forward version of the GT relation, which was used extensively in
refs.[10-12], can be found from eq.(3.2) by using the Ward identities
(2.5) for the propagators together with (2.13) for $\C_{V_{\m5}^a NN}$.
This allows us to write, for all $k$,
$$\eqalignno{
2m_N G_A^a(k^2) + k^2 G_P^a(k^2) = &- \Bigl(2n_f \d_{a0} W_{\o\o} + M_{ac}
W_{S_5^c\o}\Bigr) \hat\C_{QNN} \cr
&- \Bigl(2n_f \d_{a0} W_{\o S_5^b} + M_{ac} W_{S_5^c S_5^b}\Bigr)
\hat\C_{\phi_5^b NN}
&(3.5) \cr }
$$
$G_P^a(k^2)$ is the pseudoscalar form factor in the non-forward matrix
element $\langle p,s|J_{\m5}^a|p,s\rangle$, and again has no $1/k^2$ pole
for the reasons given above. This expression clearly reduces to eq.(3.4) on
using the zero-momentum Ward identities (2.11) for the propagators.

The remaining step to convert eq.(3.4) into the useful form of the GT
relations is to normalise the field $\phi_5^a$ appropriately. Clearly,
eq.(3.4) is independent of the normalisation. However, with a suitable
choice, the vertices can be made both RG invariant and essentially
identical to the physical Goldstone boson couplings $g_{\pi NN}$ etc. To
achieve this, we define normalised fields $$
\eta^a = B_{ab} \phi_5^b
\eqno(3.6)
$$
where $B$ is a constant matrix\footnote{$\eightpoint{}^{(2)}$}{\eightpoint
\noindent From the Ward identity (in matrix notation) $$
ik_\m \C_{V_{\m5} \eta} + \Phi B^T ~\C_{\eta\eta} - M B^{-1} = 0 $$
and writing
$$
\C_{V_{\m5} \eta} = i k_\m f(k^2)
$$
we find
$$
\C_{\eta\eta} = k^2 (\Phi B^T)^{-1} f~
+~ (\Phi B^T)^{-1} M \Phi (B\Phi)^{-1}
$$
The normalisation condition (3.7) can therefore be applied while keeping
$B$ in eq.(3.6) a constant. This determines (see Appendix B for the
relation to PCAC)
$$
f(0) = \Phi B^T
$$
However, it is {\it not} possible, as in the chiral limit case[11], to
impose a normalisation valid for all $k$, such as $\C_{\eta^a \eta^b} = k^2
\d_{ab} - (m_{\eta}^2)_{ab}$, by allowing $B$ to be a function of $k^2$.}
such that 
$$
{d\over dk^2} \C_{\eta^a \eta^b} \Big|_{k=0} = \d_{ab} 
\eqno(3.7)
$$
This condition ensures that the fields $\eta^a$ have unit coupling to the
Goldstone bosons. 
The case of the singlet $\eta^0$ is of course special, since it is only
after mixing with the topological field $Q$ (and then flavour mixing)
that it becomes the physical $\eta'$. The intricacies of this are discussed 
fully in ref.[11]. Indeed,
this is why it is most convenient to impose the normalisation condition as
above on the matrix of 2-point vertices $\C_{\eta^a \eta^b}$, which is the
inverse of the pseudoscalar propagator matrix, since this most simply
characterises the $\eta^0$ before mixing with $Q$.

The proof that the vertices $\C_{\eta^a NN}$ defined with the fields
normalised according to eq.(3.7) are RG invariant now goes through in the
same way as shown in the chiral limit in ref.[11]. In Appendix C, we 
summarise some of the main results for the extension to non-zero
quark masses. Ref.[11] also contains a careful
description of how our formalism is related to standard current algebra
(PCAC). A simple illustration of this is included in Appendix B.

Re-expressing eq.(3.4) in terms of the properly normalised vertices, we have
$$
2m_N G_A^a =  \Phi_{ac} B_{cb}^T ~\hat\C_{\eta^b NN}\big|_{k=0} \eqno(3.8)
$$
where $B$ is to be determined from the 2-point vertex condition 
$$
{d\over dk^2} \C_{\phi_5^a \phi_5^b}\big|_{k=0} = B_{ac}^T ~{d\over
dk^2}\C_{\eta^c \eta^d}\big|_{k=0} ~B_{db} = B_{ac}^T B_{cb}
\eqno(3.9)
$$

The straightforward approach to finding $\C_{\phi_5^a \phi_5^b}$ is to
write it as one component of the inverse of the propagator matrix 
$W_{{\cal S} {\cal S}}$ (with ${\cal S} = \{\o,S_5^a\}$). 
This was the approach used
in refs.[10,11] to relate $\C_{\phi_5^0 \phi_5^0}$ to the topological
susceptibility $W_{\o\o}$. However, inverting this matrix in the
multi-flavour case is cumbersome, so here we use an alternative approach.

First, we use the Ward identities (2.14) to express $\C_{\phi_5^a
\phi_5^b}$ in terms of $\C_{V_{\m5}^a V_{\n5}^b}$:
$$\eqalignno{
k_\m k_\n \C_{V_{\m5}^a V_{\n5}^b} ~&=~ ik_\n \Phi_{ac} ~\C_{\phi_5^c
V_{\n5}^b} \cr
&{}\cr
&= ~\Phi_{ac} ~\C_{\phi_5^c \phi_5^d} ~\Phi_{db}~ - ~M_{ac} \Phi_{cb}
&(3.10) \cr }
$$
Then, using the general identity (A.9) for partial Legendre transforms, we
relate $\C_{V_{\m5}^a V_{\n5}^b}$ to the 2-current propagator $W_{V_{\m5}^a
V_{\n5}^b}$:
$$
\C_{V_{\m5}^a V_{\n5}^b} ~=~ W_{V_{\m5}^a V_{\n5}^b}~ - ~W_{V_{\m5}^a {\cal
S}} ~W_{{\cal S} {\cal T}}^{-1}~ W_{{\cal T} V_{\n5}^b}
\eqno(3.11)
$$
where ${\cal S}$ and ${\cal T}$ represent $\{\o,S_5^a\}$. Combining with
eqs.(2.9) and (2.15) finally yields $$
\Phi_{ac}~ \C_{\phi_5^c \phi_5^d}~ \Phi_{db} ~= ~ W_{S_D^a S_D^b} ~+~ k_\m
W_{V_{\m5}^a {\cal S}} ~W_{{\cal S} {\cal T}}^{-1}~ W_{{\cal T} V_{\n5}^b}
k_\n
\eqno(3.12)
$$
and so, in matrix notation,
$$
\Phi B^T B \Phi ~=~ {d\over dk^2} W_{S_D S_D}\big|_{k=0}~ +~ {d\over
dk^2}\bigl(k_\m W_{V_{\m5} {\cal S}} ~W_{{\cal S} {\cal T}}^{-1}~ W_{{\cal
T} V_{\n5}} k_\n\bigr)\big|_{k=0} \eqno(3.13)
$$

The argument is almost complete. $\Phi B^T$ is precisely the combination we
need for the GT relations (3.8), and is related by eq.(3.13) directly to
the first moment of the 2-point correlation function (2.10) for the
divergences of the currents. This generalises the relation with the
topological susceptibility found in refs.[10,11].

It remains to show that the final term in eq.(3.13) vanishes. The first and
last factors are of $O(k^2)$, so this will contribute zero unless there is
a $1/k^2$ pole in the inverse pseudoscalar propagator matrix $W_{{\cal S}
{\cal T}}^{-1}$. As already mentioned, there are no $1/k^2$ poles in the
propagators themselves, so all we need show is that the determinant
$\D(k^2)$ of this propagator matrix is non-vanishing at $k=0$. This follows
from the formula
$$
\D(0) = W_{\o\o}(0)~ ({\rm det} M)^{-1}~ {\rm det} \Phi \eqno(3.14)
$$
since ${\rm det} M$ and ${\rm det} \Phi$ are non-zero (eqs.(2.8)) and
$W_{\o\o}$ is non-vanishing away from the chiral limit. An elegant proof of
eq.(3.14), based on the zero-momentum Ward identities for $W_{{\cal S}
{\cal T}}$ is as follows. Define
$$
\hat M = \left(\matrix{{\bf 1} &0 \cr 0 &M \cr}\right) ~~~~~~~~~~~~~
W = \left(\matrix{
W_{\o\o}&W_{\o S_5^b} \cr W_{S_5^a \o} &W_{S_5^a S_5^b}}\right) \eqno(3.15)
$$
with $\D = {\rm det} W$. Then,
$$\eqalignno{
{\rm det} M ~~\D &= {\rm det} \hat M ~{\rm det} W \cr &{}\cr
&= \left|\matrix{
W_{\o\o}&W_{\o S_5^b} \cr M_{ac} W_{S_5^c \o} &M_{ac} W_{S_5^c
S_5^b}}\right| &(3.16) \cr }
$$
Using the zero-momentum Ward identities (2.11) and, in the case of the
$a=0$ row taking a linear combination with the first ($\o$) row, the
determinant simplifies, leaving
$$\eqalignno{
{\rm det} M ~~\D &= \left|\matrix{
W_{\o\o}&W_{\o S_5^b} \cr 0 &-\Phi_{ab}}\right| \cr &{}\cr
&= W_{\o\o} ~{\rm det} \Phi
&(3.17) \cr }
$$
as required.

\vskip0.2cm
This completes the proof of the GT relations. To summarise, we have shown
that the flavour singlet and non-singlet axial charges are given by the
single, unified relation 
$$
2m_N G_A^a ~=~  F_{ab} ~\hat \C_{\eta^b NN}\big|_{k=0} \eqno(3.18)
$$
where $F \equiv \Phi B^T$ is determined from 
$$
FF^T = {d\over dk^2} W_{S_D S_D}\big|_{k=0} \eqno(3.19)
$$
that is,
$$
F_{ac} F_{cb}^T = \lim_{k=0} {d\over dk^2}~ i \int dx~e^{ikx}~\langle 0|
T~\pl^\m J_{\m5}^a(x)~\pl^\n J_{\n5}^b(0)|0\rangle
\eqno(3.20)
$$

In current algebra terms, the matrix $F$ determined from eq.(3.20) can be
identified, subject to the standard PCAC (pole dominance) approximation
described in Appendix B, with the pseudo-Goldstone boson decay constants. 
Under the same PCAC assumptions, the vertices can be identified with the 
low-energy meson-nucleon coupling constants,
i.e. $\hat\C_{\eta^3 NN}|_{k=0} = g_{\pi NN}$, etc.  It should be emphasised, 
however, that the relations (3.18) and (3.20) are {\it exact} -- they are 
neither dependent on this interpretation nor make any PCAC assumption or 
approximation.

Eq.(3.18) is therefore the natural generalisation of the $U(1)$ GT relation
proved in refs.[10,11]. There we showed that, in the chiral limit, 
the singlet axial charge is given by 
$$
2m_N G_A^0 = 2n_f~ \sqrt{\chi'(0)}~ \hat \C_{\eta^0 NN}\big|_{k=0} \eqno(3.21)
$$
where $\chi'(0)$ is the first moment of the topological susceptibility, viz. 
$$
\chi'(0) = \lim_{k=0} {d\over dk^2}~ i\int dx~e^{ikx}
\langle0|T~Q(x)~Q(0)|0\rangle
\eqno(3.22)
$$
The new formula extends this to include mixing with the flavour non-singlet
sector, introducing a matrix structure which replaces the simple square
root in eq.(3.21) and generalising the fields in the correlation function
to the entire divergence of the current including mass terms as well as the
anomaly $Q$.

\vskip0.7cm
\noindent{\bf 3.1~~ Flavour mixing}
\vskip0.5cm
The remaining theoretical question related to the unified GT relation
(3.18) concerns $SU(3)$ flavour mixing and the extent to which eq.(3.20)
determines $F$.

In the context of PCAC or chiral perturbation theory (incorporating
the $1/N_c$ expansion to allow the inclusion of the flavour singlet
$\eta'$ (see ref.[23] and references therein)), the problem of 
$\eta - \eta'$ mixing is currently receiving renewed attention[23,24].
According to Kaiser and Leutwyler[23], the decay constant matrix in the
$a=0,8$, $~\eta-\eta'$ sector is required to contain 4 parameters,
viz.
$$
\left(\matrix{f_{0\eta'} & f_{0\eta} \cr
f_{8\eta'} & f_{8\eta} \cr}\right) ~=~
\left(\matrix{f_0\cos\th_0 &-f_0\sin\th_0\cr
f_8\sin\th_8 &f_8\cos\th_8 \cr}\right)
\eqno(3.23)
$$
with $\th_0 \neq \th_8$, in contrast to the previous conventional
analysis which assumed $\th_0 = \th_8$. Low-energy theorems[23]
yield $f_0$, $f_8$ and $\sin(\th_0 - \th_8)$ easily, from the
diagonal and off-diagonal combinations $\sum_{P=\eta',\eta}
f_{0P}f_{0P}$, $\sum_P f_{8P}f_{8P}$ and  $\sum_P f_{0P}f_{8P}$ 
respectively, but the remaining combination of the mixing angles
is harder to identify.

Returning to our result (3.18), (3.20), we find a similar situation.
Clearly the relation (3.20) can only determine 3 parameters in $F_{ab}$
(in the a,b = 0,8 sector). However, in general our analysis 
requires $F_{ab}$ to be characterised by 4 parameters, including a
mixing angle undetermined by eq.(3.20).

To see this in more detail, consider the defining equation (3.9) 
for $B$, re-expressed in terms of $F$ itself:
$$
\Phi_{ac}{d\over dk^2} \C_{\phi_5^c \phi_5^d}\big|_{k=0} \Phi_{db}
= F_{ac} ~{d\over dk^2}\C_{\eta^c \eta^d}\big|_{k=0} ~F^T_{db} 
= F_{ac} F^T_{cb}
\eqno(3.24)
$$ 
Since the l.h.s. is symmetric, it can be diagonalised by an 
orthogonal matrix $R$, i.e.
$$
\Phi_{ac}{d\over dk^2} \C_{\phi_5^c \phi_5^d}\big|_{k=0} \Phi_{db}   
= R^T D^2 R
\eqno(3.25) 
$$
where $D$ is diagonal. But since the r.h.s. of eq.(3.24) is unchanged
if $F$ is right-multiplied by an independent orthogonal matrix $O^T$, we 
find the general solution
$$
F = R^T D O
\eqno(3.26)
$$
A convenient parametrisation for $F$ is therefore
$$
F = f 
\left(\matrix{\cos\theta & -\sin\theta \cr \sin\theta &\cos\theta}\right)
\left(\matrix{\sqrt6 s & 0 \cr 0 & 1}\right)
\left(\matrix{\cos\phi & \sin\phi \cr -\sin\phi &\cos\phi}\right)
\eqno(3.27)
$$
The four parameters are interpreted as an overall scale $f$ which in
the PCAC approximation and the limit of exact $SU(3)$ becomes the
pion decay constant, an OZI breaking ($s\neq 1$) parameter, an 
angle $\theta$ characterising $SU(3)$ breaking, and finally a 
mixing angle $\phi$ for the $\eta^a$ fields.

We can now check whether any special cases of this general parametrisation
are consistent with the RGE (C.10) for $F_{ab}$.
Clearly there is an exact $SU(3)$ limit in which both $\theta$ and $\phi$
are zero, with the RGE being satisfied by $\DD s = \c s$, $\DD f = 0$.
The case $\theta = 0$, $\phi \neq 0$ is also RG consistent, with the
same solution. (This corresponds to the case $\th_0 = \th_8$ in the
parametrisation (3.23).)
However, the RGE cannot be satisfied with $\theta \neq 0$ but $\phi = 0$.
In the general $SU(3)$ breaking case, therefore, $F_{ab}$ must 
depend on a mixing angle $\phi$ which cannot be determined from
the condition (3.20).

The consequence of this is that in order to use the full unified
GT relation (3.18) to make predictions that include the effect of $SU(3)$
flavour mixing, we would need a further (linear) condition on $F$ beyond 
eq.(3.20).

Without such an extra condition, we have to neglect flavour mixing.
This will produce an uncertainty in our final predictions which we can
estimate by defining an $SU(3)$ breaking parameter from the 
condensates (2.7), viz. $t = \sqrt6 \langle\phi^8\rangle/
\langle\phi^0\rangle$.  
(The $\sqrt6$ factors here and in the definition of $s$ arise because 
of the different normalisation of the singlet generator:
${\rm tr} T^0 T^0 = n_f$ whereas ${\rm tr} T^i T^j = {1\over2}\d^{ij}$
for $i,j = 1, \ldots 8$.) Using standard values (6.16) for the condensates,
$t \simeq 0.16$. We therefore expect the $SU(3)$ breaking angle
$\sin \theta$ to be of $O(t)$ and so omitting flavour mixing
effects will produce an uncertainty of $O(10-20\%)$ in our final results.

However, when we evaluate the 
correlation functions (3.20) using QCD spectral sum rules (section 6),
it is in any case difficult to do better, since to incorporate flavour
mixing we would have to saturate the spectral functions (below $t_c$)
with the two states $\eta'$ and $\eta$, for both the flavour diagonal
and off-diagonal correlators. This would greatly complicate the analysis
without significantly improving the accuracy of the final results,
so we do not attempt it here. Consequently, our final predictions,
presented in section 7, are subject to the approximation of
neglecting $SU(3)$ flavour mixing.

\vfill\eject

\noindent {\bf 4. GT Relations and the Sum Rule for the First Moment of
$g_1^p$} \vskip0.5cm
As already quoted in the introduction, the first moment of the polarised 
proton structure function $g_1^p$ satisfies the following sum rule: 
$$\eqalignno{
\Gamma^p_1(Q^2) &\equiv
\int_0^1 dx~ g_1^p(x;Q^2) \cr
&= {1\over12} C_1^{\rm NS}\aQ~ \Bigl( a^3 + {1\over3} a^8 \Bigr) +
{1\over9} C_1^{\rm S}\aQ~ a^0(Q^2) &(4.1) \cr}
$$
where the $a^a$ are the axial charges occurring in the GT relations we have
just derived:
$$
G_A^3 = {1\over2}a^3
~~~~~~~~~~~~~~~~~~
G_A^8 = {1\over2\sqrt3}a^8
~~~~~~~~~~~~~~~~~~
G_A^0(Q^2) = a^0(Q^2)
\eqno(4.2)
$$
The RG scale ($Q^2$) dependence of $a^0(Q^2)$, due to the anomalous
dimension of the singlet axial current, is explicitly displayed.

This sum rule has been analysed using the composite operator
propagator-vertex approach to DIS in ref.[12]. Working in
the chiral limit, it was shown that
$$
\C_1^p{}_{singlet} = {1\over9} {1\over 2m_N} 2n_f~
C_1^S\bigl(\a_s(Q^2)\bigr)~ \sqrt{\chi'(0)}\big|_{Q^2}~ \hat\C_{\eta^0NN}
\eqno(4.3)
$$
Following refs.[10-12], we now assume that the vertices are well
approximated by their OZI values. This is the key assumption that
allows us to make a quantitative prediction for $\C_1^p$ on the basis of a
calculation of the topological susceptibility alone. The RG invariance of
the vertices is a necessary condition for this assumption to be
reasonable. Further phenomenological evidence from $U_A(1)$ current
algebra supporting this conjecture is discussed in refs.[10-12].

We therefore assume that $\hat\C_{\eta^0 NN}$ satisfies the OZI rule, 
i.e. $\hat\C_{\eta^0 NN} = \sqrt2 ~\hat\C_{\eta^8 NN}$, 
so that all the OZI breaking in $\C_1^p{}_{singlet}$ resides in the 
topological susceptibility $\sqrt{\chi'(0)}$. Comparing
with the standard OZI relation for $a^8$, we then find that[10,11]
$$
{a^0(Q^2)\over a^8} ~=~ {\sqrt6 \over f_\pi}~
\sqrt{\chi'(0)}\big|_{Q^2}
\eqno(4.4)
$$
This leads to the following prediction[12] for the `proton spin'
sum rule:
$$\eqalignno{
a^0(Q^2=10\GV^2) &= 0.35 \pm 0.05 \cr
\C_1^p(Q^2=10\GV^2) &= 0.143 \pm 0.005
&(4.5) \cr }
$$
based on our original derivation of $\chi'(0)$ using QCD spectral 
sum rules[12]: 
$$
\sqrt{\chi'(0)}\big|_{Q^2=10\GV^2} = (23.2 \pm 2.4) ~{\rm MeV} 
\eqno(4.6)
$$

This is to be compared with the OZI prediction $a^0\big|_{OZI} = a^8
= 0.58 \pm 0.03$ and with the current experimental data from the SMC
collaboration[25]:
$$
\C_1^p (Q^2=10\GV^2)\big|_{(x>0.003)} \equiv \int_{0.003}^1 dx~g_1^p(x;Q^2)
= 0.141 \pm 0.012
\eqno(4.7a)
$$
The result for the entire first moment depends on how the extrapolation to
the unmeasured small $x$ region $x<0.003$ is performed. Using a simple
Regge fit, SMC find $\C_1^p = 0.142 \pm 0.017$ from which they deduce
$a^0= 0.34 \pm 0.17$, while using a small $x$ fit using perturbative QCD 
evolution of the parton distributions[4,26] they find 
$\C_1^p = 0.130 \pm 0.017$ and
$a^0 = 0.22 \pm 0.17$ ~(all at $Q^2 = 10\GV^2$). 

More recently, SMC have published a further analysis[27] of their data,
this time quoting a slightly lower number for the integral over
the measured region of $x$:
$$
\C_1^p (Q^2=10\GV^2)\big|_{(x>0.003)} \equiv \int_{0.003}^1 dx~g_1^p(x;Q^2)
= 0.133 \pm 0.009
\eqno(4.7b)
$$

Whatever the ultimate resolution of the small $x$ extrapolation, it is
encouraging that our prediction (4.5) is in the region now favoured by the
data. We would therefore like to test the precision of our result further.
An assumption in making this prediction was that the value of $a^0$ is
smooth in the quark masses, so that the chiral limit will be a good
approximation, correct up to the usual order of soft $SU(3)$ breaking
in ratios of decay constants. The extended version of the GT relations 
derived in section 3 allow us to test this.

Using the new form (3.18) of the GT relations, we can immediately rewrite
the first moment sum rule for $g_1^p$ in the following compact form:
$$
\eqalignno{
\C_1^p(Q^2) \equiv \int_0^1 dx~g_1^p(x,Q^2) &={1\over9} {1\over2m_N}~
C_1^a\bigl(\a_s(Q^2)\bigr) ~G_A^a \cr &{}\cr
&= {1\over9} {1\over2m_N}~ C_1^a\bigl(\a_s(Q^2)\bigr) ~F_{ab} ~
\hat\C_{\eta^b NN} &(4.8) \cr }
$$
where we have defined the numerically rescaled Wilson coefficients $C_1^0 =
C_1^S$ and $C_1^3 = \sqrt3 C_1^8 = {3\over2}C_1^{NS}$. All the vertices
$\C_{\eta^a NN}$ are RG scale invariant. Apart from the running coupling in
the Wilson coefficients, the only $Q^2$ scale dependence is contained in
the singlet components of $F_{ab}$, according to the RG equation (C.10).

Following the approach of refs.[10-12], we again assume that these 
RG-invariant vertices are all well approximated by their OZI values. 
Isospin invariance ensures that the off-diagonal terms in the 2-current 
correlation functions involving the triplet index 3 vanish, so we need 
consider only:
$$
FF^T = \lim_{k=0} ~{d\over dk^2}~~i
\left(\matrix{
\langle 0|\pl^\m J_{\m5}^0~\pl^\n J_{\n5}^0|0\rangle &0
&\langle 0|\pl^\m J_{\m5}^0~\pl^\n J_{\n5}^8|0\rangle \cr 0
&\langle 0|\pl^\m J_{\m5}^3~\pl^\n J_{\n5}^3|0\rangle &0 \cr
\langle 0|\pl^\m J_{\m5}^8~\pl^\n J_{\n5}^0|0\rangle &0
&\langle 0|\pl^\m J_{\m5}^8~\pl^\n J_{\n5}^8|0\rangle \cr} \right)
\eqno(4.9)
$$
In section 6, we evaluate these correlators (in the $0,8$ sector)
using spectral sum rules. As explained in the last section, we restrict
ourselves to single-particle saturation of the appropriate spectral
functions, keeping only the $\eta'$ contribution in the flavour singlet
correlator and $\eta$ in the octet. We do not evaluate the off-diagonal
correlator, which is expected to be small due to the approximate 
cancellation between the decay constants for the $\eta'$ and 
$\eta$ states in the spectral function. 

In this approximation of consistently neglecting the $SU(3)$ flavour
mixing, and using the OZI relation $\hat\Gamma_{\eta^0 NN}
= \sqrt2 ~\hat\Gamma_{\eta^8 NN}$, we then identify the `proton spin'
suppression from eq.(4.8) as:
$$
{a^0(Q^2) \over a^8}  ~\simeq~
{1\over \sqrt6}~ {F_{00} \over F_{88}}
\eqno(4.10)
$$
where
$$
F_{00} ~\simeq~ \sqrt{\lim_{k=0} ~{d\over dk^2}~
i\langle 0|\pl^\m J_{\m5}^0~\pl^\n J_{\n5}^0|0\rangle }
\eqno(4.11)
$$
and 
$$
F_{88} ~\simeq~ \sqrt{\lim_{k=0} ~{d\over dk^2}~
i\langle 0|\pl^\m J_{\m5}^8~\pl^\n J_{\n5}^8|0\rangle }
\eqno(4.12)
$$

\vskip1cm

\noindent {\bf 5. Quark Mass Dependence} 
\vskip0.5cm
Before presenting the spectral sum rule analysis in section 6, we give 
here some analytic arguments based on the Ward identities which suggest
that the first moment of the correlation functions (4.9) will have only 
a weak dependence on the quark masses. The basic reason for this
is the identification,
up to the standard PCAC pole-dominance assumptions, of $F$ with a decay
constant. The explicit dependence on the pseudo-Goldstone boson masses
which is present in $W_{S_D^a S_D^b}$ drops out in the first moment.

To see this another way, consider the Ward identity (2.9). Taking the first
moment, we see that the explicit dependence on the quark masses $m$
vanishes, leaving
$$
FF^T ~~=~~
{d\over dk^2} W_{S_D^a S_D^b}\big|_{k=0} ~~=~~ {d\over dk^2} \Bigl( k^\m
k^\n W_{V_{\m5}^a V_{\n5}^b} \Bigr)\Big|_{k=0} \eqno(5.1)
$$
The 2-current correlation function can be parametrised as $$\eqalignno{
W_{V_{\m5}^a V_{\n5}^b} ~~&=~~
i \int dx~ e^{ik.x}~ \langle0|T~ J_{\m5}^a(x)~J_{\n5}^b(0)|0\rangle \cr
&=~~\Pi_T^{ab}(k^2) \Bigl(g_{\m\n} - {k_\m k_\n\over k^2} \Bigr) ~~+~~
\Pi_L^{ab}(k^2) {k_\m k_\n\over k^2}
&(5.2) \cr }
$$
where $\Pi_T^{ab}(k^2)$, $\Pi_L^{ab}(k^2)$ are dynamical functions which
are not determined by the chiral Ward identities. (They are of course
related by eqs.(2.5) to other correlation functions.) While
$\Pi_L^{ab}(k^2)$ will have poles corresponding to the pseudo-Goldstone
bosons coupling to the currents, $\Pi_T^{ab}(k^2)$ is a pole-free
function (at least for momenta below the masses of the pseudovector
resonances). $\Pi_T^{ab}(0)$ is therefore not expected to have a
substantial dependence on the pseudoscalar masses. The absence of a
massless pseudoscalar pole in (5.2) requires $\Pi_L^{ab}(0) = \Pi_T^{ab}(0)$,
so clearly
$$
{d\over dk^2} W_{S_D^a S_D^b}\big|_{k=0} ~~=~~ \Pi_L^{ab}(0)
~~=~~\Pi_T^{ab}(0) \eqno(5.3)
$$

\vskip0.2cm

Although we do not expect a strong mass dependence of the correlation
functions (5.1), the individual correlators in the decomposition
$$\eqalignno{
{d\over dk^2} W_{S_D^a S_D^b}\big|_{k=0} ~~=~~ &4n_f^2 \d_{a0}\d_{b0}
{d\over dk^2}W_{\o\o}\Big|_{k=0} ~~+~~ 2n_f \d_{a0} M_{bd} {d\over
dk^2}W_{\o S_5^d}\Big|_{k=0} \cr &+ 2n_f \d_{b0} M_{ac} {d\over
dk^2}W_{S_5^c \o}\Big|_{k=0} ~~+~~ M_{ac} M_{bd} {d\over dk^2}W_{S_5^c
S_5^d}\Big|_{k=0} &(5.4) \cr}
$$
certainly do have an explicit dependence on $m$. This of course cancels in
the sum, although the pattern of cancellations is very intricate.

\vskip0.2cm
To illustrate all this, it is instructive to write a simple effective
action $\C[Q,\phi_5^a]$ which encodes the information in the zero-momentum
chiral Ward identities, and use this to derive explicit expressions for the
correlation functions in eq.(5.4).

The zero-momentum Ward identities are
$$
\Phi_{ac} \C_{Q\phi_5^c} \Big|_{k=0} = 2n_f \d_{a0} ~~~~~~~~~~~~ \Phi_{ac}
\C_{\phi_5^c \phi_5^b} \Big|_{k=0} = M_{ab} \eqno(5.5)
$$
The simplest effective action compatible with these identities is $$
\C[Q,\phi_5^a] ~~=~~\int dx~~\biggl[{1\over2a} Q^2 ~+~ 2n_f Q
\Phi_{0a}^{-1} \phi_5^a ~+~
{1\over2} \phi_5 \Phi^{-1} f \bigl(-\pl^2 -\m^2\bigr) f \Phi^{-1} \phi_5
\biggr]
\eqno(5.6)
$$
The final term is written in matrix notation. $f_{ab}$ and $\m_{ab}^2$ are
matrices, $a$ and $f_{ab}$ are constant, and $\m_{ab}^2$ is {\it defined} by
$$
f_{ac} ~\m_{cd}^2 ~f_{db} ~=~ -M_{ac} \Phi_{cb} \eqno(5.7)
$$
$\m^2$ is the pseudo-Goldstone boson mass matrix in the OZI limit of QCD,
i.e.~neglecting the coupling to the anomaly $Q$.

It is important to realise that the effective action (5.6) is only an
approximation, where the simplest choice of kinetic terms for the fields
$\phi_5^a$ has been made. This corresponds to the pole dominance
approximation in standard PCAC (see Appendix B). In this approximation, as
we now show, there is strictly no $m$ dependence in (5.1).

The second derivatives of the effective action are $$
\left(\matrix{\C_{QQ} &\C_{Q\phi_5^b}\cr \C_{\phi_5^a Q} &\C_{\phi_5^a
\phi_5^b} \cr}\right) ~~=~~ \left(\matrix{a^{-1} &2n_f \Phi_{0b}^{-1} \cr
2n_f \Phi_{a0}^{-1} & \Phi^{-1} f \bigl(k^2-\m^2\bigr) f \Phi^{-1}\cr
}\right) \eqno(5.8)
$$
The correlation functions are found by inverting this matrix. We find
$$\eqalignno{
W_{\o\o} ~~&=~~ -a~ \tilde\D^{-1} \cr
W_{\o S_5^b} ~~&=~~ 2n_f a~ \D_{0d}^{-1}~ \Phi_{db} \cr 
W_{S_5^a \o} ~~&=~~ 2n_f a~ \Phi_{ac}~
\Bigl(f\bigl(k^2-\m^2\bigr)f\Bigr)_{c0}^{-1} \tilde\D^{-1} \cr
W_{S_5^a S_5^b} ~~&=~~ - \Phi_{ac}~ \D_{cd}^{-1}~ \Phi_{db} &(5.9) \cr }
$$
where
$$
\tilde \D ~=~ 1 - 4n_f^2 a \Bigl(f\bigl(k^2-\m^2\bigr)f\Bigr)_{00}^{-1}
\eqno(5.10)
$$
and
$$
\D ~=~ f\bigl(k^2-\m^2\bigr)f - 4n_f^2 a~ \hat I \eqno(5.11)
$$
with $\hat I = \d_{a0}\d_{b0}$. Notice the highly non-trivial $\m^2$
dependence in all these correlators.\footnote{$\eightpoint {}^{(3)}$}
{\eightpoint
\noindent The expressions (5.9) are (in a different notation) identical to
those derived in ref.[28] from an effective action or diagrammatic
resummation including effects beyond leading order in the $1/N_c$
expansion.}

A simple calculation now confirms that (recalling $D^a = 2n_f \d_{a0} Q +
M_{ac} \phi_5^c$) the zero-momentum Ward identity
$$
W_{S_D^a S_D^b}\Big|_{k=0} ~~=~~ -M_{ac}\Phi_{cb} \eqno(5.12)
$$
is satisfied. A slightly trickier calculation also shows that the $M$ and
$\m^2$ terms cancel completely in the first moment and we find
$$
{d\over dk^2} W_{S_D^a S_D^b}\Big|_{k=0} ~~=~~ f_{ac} f_{cb} \eqno(5.13)
$$
In fact, this is required by the key identity (3.9), which shows $$
{d\over dk^2} W_{S_D^a S_D^b}\Big|_{k=0} ~~=~~ \Phi_{ac} {d\over dk^2}
\C_{\phi_5^c \phi_5^d}\Big|_{k=0} \Phi_{db} \eqno(5.14)
$$
The first moment of the correlation function $W_{S_D^a S_D^b}$ is therefore
given by the decay constants in the simple effective action (5.6).

Notice however that $f_{ab}$ is only constrained to be a constant by the
pole-dominance approximation, {\it not} by the chiral Ward identities. In
general, the $f_{ab}$ appearing in eq.(5.8) could be functions of momentum,
i.e. $f_{ab}(k^2)$. In that case, (5.13) is replaced by $$\eqalignno{
{d\over dk^2} W_{S_D^a S_D^b}\Big|_{k=0} ~~&=~~ {d\over dk^2}
\Bigl(f\bigl(k^2-\m^2\bigr)f\Bigr)\Big|_{k=0} \cr &=~~f^2(0) -
f(0)\m^2f'(0) - f'(0)\m^2f(0) &(5.15) \cr }
$$
We see therefore that ${d\over dk^2} W_{S_D^a S_D^b}\Big|_{k=0}$ {\it can}
have a dependence on the quark masses, but only proportional to the
derivative $f'(0)$. If, as is consistent with the success of PCAC, the
pole-free dynamical function $f(k^2)$ is slowly-varying in the
small-momentum region, then this dependence is relatively weak.

\vskip0.2cm
The conclusion of all this is that $FF^T = {d\over dk^2} W_{S_D^a
S_D^b}\Big|_{k=0}$ has only a weak residual quark mass dependence
proportional to the derivative of a slowly varying dynamical function. The
strong, explicit dependence on the quark masses cancels amongst the four
individual correlation functions in eq.(5.4).

\vskip0.2cm
The same pattern of cancellations is observed if we now write spectral sum
rules for ${d\over dk^2} W_{S_D^a S_D^b}\Big|_{k=0}$ using simply the 
pole-dominance approximation for the propagators given in eq.(5.9). 
At first sight, we might therefore expect this pattern to be reproduced 
in the full QCD spectral sum rules described in the next section. 
In fact that is not so.
Of course the sum rules go beyond the pole-dominance approximation, but
in addition, for Green functions such as those needed here, which satisfy 
dispersion relations requiring subtractions, even the sign of the corrections to
pole-dominance is not simply determined on general grounds. Nevertheless,
as we shall see, the full spectral sum rules do confirm the picture
outlined here by showing numerically the relative insensitivity of ${d\over
dk^2} W_{S_D^a S_D^b}$ to the strange quark mass.

\vfill\eject

\noindent {\bf 6. QCD Spectral Sum Rules and Current Correlation Functions}

\vskip0.5cm
First, we derive the QCD spectral sum rules (for a review, see e.g.~ref.[29])
for the flavour-singlet current correlation function 
$$
\psi_5(k^2) ~~=~~\Bigl({1\over 2n_f}\Bigr)^2 i \int d^4x~e^{ikx} \langle
0|T~ \pl^\mu J_{\mu5}^0(x) ~
\pl^\nu J_{\nu 5}^0(0)~|0\rangle
\eqno(6.1)
$$
Recall the anomalous current conservation equation (c.f.~eq.(2.2)) is 
$$
\pl^\mu J_{\mu 5}^0  = D^0 =\sum_{q=u,d,s} 2m_q\bar q\c_5 q + 2n_fQ
\eqno(6.2)
$$

From its analyticity properties and asymptotic behaviour, the correlator
obeys the subtracted dispersion relations 
$$
{1 \over k^2}\Bigl[\psi_5(k^2)-\psi_5(0)\Bigr]~=~ \int_0^\infty {dt \over
t} {1 \over (t-k^2-i\e)}~ {1\over\pi}{\rm Im}\psi_5(t)
\eqno(6.3)
$$
and
$$
{1\over k^4}\Bigl[\psi_5(k^2)-\psi_5(0)-k^2\psi'_5(0)\Bigr] ~=~
\int_0^\infty {dt \over t^2} {1 \over (t-k^2-i\e)}~ {1 \over \pi} {\rm
Im}\psi_5(t)
\eqno(6.4)
$$
From the leading large $K^2\equiv -k^2>0$ behaviour of the
correlator, which is $K^4 \log (K^2/\mu^2)$, one can deduce that the
corresponding derivatives 
$$
{\cal F}={d^2 \over (dK^2)^2} \Bigl({\psi_5 \over K^2}\Bigr)~~~~~~~~~~~~
{\cal G}=-{d \over dK^2} \Bigl({\psi_5 \over K^4}\Bigr) \eqno(6.5)
$$
are superconvergent and thus obey the homogeneous RGE: 
$$
\Bigl[-{\pl\over \pl t}+\b\as{\pl\over\pl \as}-
\sum_i(1+\c_m)x_i{\pl\over\pl x_i} - 2\c \Bigr] ({\cal F;G})(t, \as, x_i)=0
\eqno(6.6)
$$
Here, $x_i\equiv m_i/\mu$ is the ratio of the renormalised quark mass with
the $\msb$-scheme subtraction scale $\mu$, and $t\equiv L/2$ where 
$$
L\equiv \log (K^2/\mu^2)
\eqno(6.7)
$$
$\b$, $\c$ and $\c_m$ are respectively the QCD $\b$-function, the anomalous
dimension for $J_{\m5}^0$ and the mass anomalous dimension. The anomalous
dimension $\c$ is $O(\alpha_s^2)$, viz. $$
\c(\a_s) = - \asp^2
\eqno(6.8)
$$
and does not contribute at the order we consider here. The coefficients of
$\b$ and $\c_m$ are
$$
\b(\as)=\sum_{i=1}\b_i\asp^i ~~~~~~~~~
\c_m(\as)=\sum_{i=1}\c_i\asp^i
\eqno(6.9)
$$
where, for three flavours[29],
$$\eqalignno{
\b_1 &= -9/2 ~~~~~~\b_2 = -8 \cr
\c_1 &=2 ~~~~~~~~~~\c_2 = 91/12 &(6.10) \cr }
$$

The expression for the running coupling
to two-loop accuracy can be parametrised as[29]: 
$$
a_s(\mu)\equiv {\bar{\alpha}_s(\mu)\over\pi} =
a_s^{(0)}\Bigl[1-a_s^{(0)}{\b_2 \over \b_1}\log\log{\mu^2\over\L^2} +
O(a_s^2)\Bigr]
\eqno(6.11)
$$
where
$$
a_s^{(0)} \equiv {1 \over -\b_1\log(\mu/\L)} \eqno(6.12)
$$
and
$\b_i$ are the coefficients of the $\b$ function given above. We shall use,
for three flavours,
$$
\L= (375 \pm 75)~\MV
\eqno(6.13)
$$
from $\t$ decay [30] and LEP[31] data.

The expression for the running quark mass in terms of the invariant mass
$\hat{m}_i$ to two-loop accuracy is \footnote{$\eightpoint
{}^{(4)}$}{\eightpoint \noindent The truncation of the series at this order
is necessary for self-consistency as the quark-quark correlator will be
used to $O(\as)$.}[32,29]: 
$$
\bar m_i(\mu) ~=~ \hat{m}_i ~\bigl(-\b_1 a_s(\mu)\bigr)^{-\c_1/\b_1}
\Bigl[1+{\b_2 \over \b_1}\Bigl({\c_1 \over \b_1}-{\c_2 \over \b_2}\Bigr)
a_s(\mu) + O(a_s^2) \Bigr]
\eqno(6.14)
$$
where $\c_i$ are the coefficients of the quark-mass anomalous dimension
given above. In this analysis, we shall retain only the strange quark mass
and neglect $m_u$ and $m_d$. We use
$$
\bar m_s(1~{\rm GeV}) \simeq (197\pm 29)~{\rm MeV} \eqno(6.15)
$$
from the $e^+e^-\rightarrow$ hadrons[33] and $\t$ decay[34] data, and
the correlated values of the invariant mass $\hat{m}_s$ and $\L$. We shall
also use[29]: $$
\langle \bar ss\rangle \simeq (0.6\sim 0.8)\langle \bar uu\rangle ~~~~~~~~
\langle \bar uu\rangle = -(0.238~\rm{GeV})^3 \eqno(6.16)
$$

\vskip0.3cm
Now, applying the inverse Laplace operator[35] 
$$
{\cal L}\equiv
\lim_{K^2,n\rightarrow\infty;~ n/K^2\equiv \t}~ (-1)^n{(K^2)^n \over
(n-1)!} {\pl^n \over(\pl K^2)^n} \eqno(6.17)
$$
to the dispersion relations (6.3) and (6.4) gives the sum rules
\footnote{$\eightpoint {}^{(5)}$}{\eightpoint \noindent The unsubtracted
sum rule, which is independent of $\psi_5(0)$ and $\psi'_5(0)$, is more
sensitive to the higher meson states, and thus is more appropriate for
studying gluonium parameters[36].}
$$\eqalignno{
\t^{-3} {\cal L}({\cal F})+\psi_5(0) &=\int_0^\infty {dt \over t}~e^{-t\t}
{1\over\pi} {\rm Im}\psi_5(t) &(6.18) \cr \t^{-2} {\cal L}({\cal
G})-\psi_5(0)\t+\psi'_5(0) &=\int_0^\infty {dt\over t^2}~e^{-t\t}
{1\over\pi} {\rm Im}\psi_5(t) &(6.19) \cr }$$
We use the usual duality ansatz for parametrising the spectral function: 
$$
{1\over\pi}{\rm Im}\psi_5(t)~\simeq~ 2m^4_{\eta'} f^2_{\eta'}
\d\bigl(t-m^2_{\eta'}\bigr)~
+~\theta (t-t_c)~``{\rm QCD}~{\rm continuum}" \eqno(6.20)
$$
where $f_{\eta'}$ is the RG non-invariant $\eta'$ `decay constant' 
(see refs.[11,37]) normalised as
$$
\langle 0~|\pl^\mu J_{\mu 5}^{(0)}|~\eta'\rangle = 2n_f ~\sqrt2~ f_{\eta'}
m^2_{\eta'}
\eqno(6.21)
$$
Then, after transferring the QCD continuum contribution to the lhs of
eqs.(6.18) and (6.19), we obtain the sum rule: 
$$
\psi'_5(0) ~=~ \t^{-2}\Bigl[-{\cal L}({\cal G}_c)+{1\over\t m^2_{\eta'}}
{\cal L}({\cal F}_c)\Bigr]
+\psi_5(0)\t\Bigl[1+{1\over\t m^2_{\eta'}}\Bigr] \eqno(6.22)
$$
where the index $c$ in ${\cal F}_c$ and ${\cal G}_c$ indicates that the QCD
continuum effect has been transferred into the QCD expression of the
correlators\footnote{$\eightpoint {}^{(6)}$}{\eightpoint \noindent In the
following, we shall omit this index for convenience of notation.}.

Analogously to ${\cal F}$ and ${\cal G}$, their Laplace transforms also
obey an homogeneous RGE [38], where the resummation of the log-terms
can be done by subtracting at $\t=1/\mu^2$ and introducing the running
coupling and masses.

At this point, it is convenient to decompose the full correlation function
(6.1) as follows
$$
\psi_5(k^2) ~\equiv~
\psi_{gg}(k^2)+2\Bigl({m_s \over n_f}\Bigr)\psi_{qg}(k^2)+ \Bigl({m_s \over
n_f}\Bigr)^2\psi_{qq}(k^2) \eqno(6.23)
$$
where
$$\eqalignno{
\psi_{gg}(k^2) \equiv \chi(k^2) &\equiv i\int d^4x~e^{ikx}\langle 0|T~
Q(x)~Q(0)|0\rangle\cr
\psi_{qg}(k^2) &\equiv i\int d^4x~e^{ikx}\langle 0|T~ Q(x)~\bar s\c_5
s(0)~|0\rangle \cr
\psi_{qq}(k^2) &\equiv i\int d^4x~e^{ikx}\langle 0|T~ \bar s\c_5 s(x) ~\bar
s\c_5 s(0)~|0\rangle &(6.24) \cr }$$
Here, we have used the expression (6.2) for the divergence of the current
and neglected the $u$ and $d$ quark masses.

For the next stage in developing the sum rules, we need the perturbative
QCD expressions for these correlators. These have been calculated in the
literature.
First, the perturbative QCD expression for the gluon-gluon correlator
$\psi_{gg}(k^2)$ has been reported in [12] and reads: 
$$
\psi_{gg}(k^2)=\psi_{gg}^{PT}+\psi_{gg}^{NP}, \eqno(6.25)
$$
where
$$\eqalignno{
\psi_{gg}^{PT}&=-\Bigl({\as \over 8\pi}\Bigr)^2{2 \over \pi^2}k^4L \biggl[
1+\asp\Bigl({1\over2}\b_1L+{83\over4} +... +6\bigl(3m_s^2 +
{\pi\over\a_s}\l^2\bigr){1\over k^2} \Bigr)\biggr] \cr
\psi_{gg}^{NP}&=-\Bigl({\as \over 16\pi^2}\Bigr) \biggl[ \Bigl(
1+{1\over2}\b_1\asp L \Bigr) \langle\as G^2\rangle +{2 \over K^2}\as
\langle gG^3\rangle \biggr] &(6.26) \cr }$$

The new corrected coefficient of the perturbative $O(\alpha_s^3)$ term
comes from the erratum in [39].
Notice that we have computed the $m_s$-dependent contribution of
$O(\alpha_s^3)$ coming from the 2-loop Feynman diagram with a quark loop
inserted on one gluon propagator. This is a new calculation that has not
previously been published in the literature.\footnote{$\eightpoint
{}^{(7)}$}{\eightpoint
\noindent We thank Alexei Pivoravov for checking this result.}

We have also included the correction due to a tachyonic gluon mass
$\lambda$, where[40]:
$$
\lambda^2 \simeq - (0.43 \pm 0.09) ~\GV^2
\eqno(6.27)
$$
in order to take into account the summation of the perturbative series (as
a phenomenological alternative to renormalons). A full discussion of the
motivation for including this term and its phenomenology is given in
ref.[40].

The non-perturbative contributions come
from [41] and [42]. Throughout the analysis, we shall use the values
of the gluon condensates [43]: $$
\langle \as G^2\rangle = (0.07\pm 0.01)~{\rm GeV}^4 \eqno(6.28)
$$
and [41]
$$
\langle g^3 G^3\rangle=(1.5\pm 0.5)~{\rm GeV}^2\langle \as G^2\rangle
\eqno(6.29)
$$
A similar value of $\langle \as G^2\rangle$ has been obtained recently from
lattice calculations [44].

The QCD expression for the quark-gluon correlator $\psi_{qg}(k^2)$ has been
evaluated in [45]. In the $\msb$-scheme, it reads: 
$$
\psi_{qg}(k^2)=\psi_{qg}^{PT}+\psi_{qg}^{NP}, \eqno(6.30)
$$
where
$$\eqalignno{
\psi_{qg}^{PT}&= \asp^2 m_s {3 \over 16\pi^2}k^2L \Bigl[L-{2 \over
3}\Bigl({11 \over 4}-3\c_E\Bigr)\Bigr] \cr \psi_{qg}^{NP}&=
-\asp^2\langle\bar ss\rangle {L \over 2}+\asp {m_s \over 8\pi} \langle \as
G^2\rangle {L \over k^2} + \asp {1 \over 2k^2}\langle g\bar s\s^{\mu\nu}
{\l_a \over 2}G^a_{\mu\nu}s \rangle &(6.31) \cr
}$$
and $\c_E = 0.5772\ldots $ is the Euler constant.

Finally, the QCD quark-quark correlator $\psi_{qq}(k^2)$ is known to order
$O(\alpha_s^3)$ for the perturbative term [46]. Including the condensates
of dimension 6 [35,29], it can be expressed as: 
$$
\psi_{qq}(K^2\equiv-k^2)=K^2\sum_{d=0}^3 {\psi_{2d}(K^2)\over K^{2d}}
\eqno(6.32)
$$
In the $\msb$-scheme, the perturbative expression of the renormalised
two-point function can be written as:
$$
\psi_0(K^2)={3 \over 8\pi^2}\sum_{i=0}\asp^i~\sum_{j=0}^{i+1}c_{ij}L^j
\eqno(6.33)
$$
where $c_{ij}$ are constant terms from the evaluation of the QCD Feynman
diagrams. To $O(\alpha_s^2)$, they read [47,45]: 
$$\eqalignno{
c_{00}&=-2 \cr
c_{10}&=-131/12 \cr
c_{20}&={1 \over 6}\Bigl[-17645/24+(353-8n_f)\zeta(3)+(511/18)n_f
+(3/4)\zeta(4)-50\zeta(5)\Bigr] \cr
c_{01}&=1 \cr
c_{11}&=17/3 \cr
c_{12}&=-{1 \over 2}\c_1 c_{01} \cr
c_{21}&=10801/144-(39/2)\zeta(3)-n_f\Bigl[65/24-(2/3)\zeta(3)\Bigr] \cr
c_{22}&=-{1 \over 4}\Bigl[c_{11}(-\b_1+2\c_1)+2\c_2 c_{01}\Bigr] \cr
c_{23}&={1 \over 12}\c_1(-\b_1+2\c_1)c_{01} &(6.34) \cr }$$
where $\zeta(n)$ are the Riemann zeta functions with $$
\zeta(2) = {\pi^2\over6} ~~~~~~~~~~~\zeta(3) = 1.202\ldots \eqno(6.35)
$$
Given the approximations and accuracy to which we are working, we only keep
the following chiral symmetry breaking and non-perturbative condensate
contributions:
$$\eqalignno{
\psi_2&={3 \over 8\pi^2}m^2_s\Bigl[2L+{4 \over 3}\asp\bigl(
-3L^2+2L-3+6\zeta(3)\bigr)\Bigr] \cr
\psi_{4}&={3 \over 8\pi^2}m^4_s\bigl(3-2L\bigr) -m_s\langle \bar ss\rangle
+ {1 \over 8\pi}\langle \as G^2\rangle \cr \psi_{6}&=-m_s\langle g\bar
s\s^{\mu\nu} {\l_a \over 2}G^a_{\mu\nu}s\rangle +{112 \over
27}\pi\rho\as\langle\bar ss\rangle^2 &(6.36) \cr }$$
The mixed condensate can be parametrised as: $$
\langle g\bar s\s^{\mu\nu} {\l_a \over 2}G^a_{\mu\nu}s\rangle =
M^2_0\langle \bar ss\rangle
\eqno(6.37)
$$
where the value of $M^2_0=(0.8\pm 0.1)~{\rm GeV}^2$ comes from the baryon
[48,49] and $B^*$-$B$ [50] sum rules and $\rho=2\sim 3$ [43,51]
indicates the deviation from the vacuum saturation estimate of the
four-quark operators.

\vskip0.3cm
To complete the sum rules, we need the Laplace tranforms of ${\cal F}$ and
${\cal G}$. These can be obtained from the renormalised QCD expressions
above with the help of the generic formulae: 
$$\eqalignno{
{\cal L}\biggl[ {1 \over K^{2n}}\biggr]&= {\t^n \over \C(n)} \cr
{\cal L}\biggl[ {L \over K^{2n}}\biggr]&= {\t^n \over
\C(n)}\Bigl[L_\t+\psi(n)\Bigr] \cr {\cal L}\biggl[ {L^2 \over
K^{2n}}\biggr]&= {\t^n \over
\c(n)}\Bigl[L^2_\t+2\psi(n)L_\t+\psi^2(n)-\psi'(n)\Bigr] \cr {\cal
L}\biggl[ {L^3 \over K^{2n}}\biggr]&={\t^n \over \C(n)}
\Bigl[L^3_\t+3\psi(n)L^2_\t+3\bigl(\psi^2(n)-\psi'(n)\bigr)L_\t+
\psi^3(n)-3\psi(n)\psi'(n)+\psi^{''}(n)\Bigr] \cr &\hfil &(6.38) \cr
}$$
where
$$\eqalignno{
L_\t&\equiv -\log{\t\mu^2} \cr
\psi(1)&=-\c_E \cr
\psi'(1)&=\zeta(2) \cr
\psi''(1)&=-2\zeta(3) \cr
\psi(n)&=\sum_{j=1}^{n-1} {1 \over j}-\c_E \cr \psi^{(k)}(n)&=(-1)^k k!
\biggl[\sum_{j=1}^{n-1} {1 \over j^{k+1}}- \zeta(k+1)\biggr] ~~~~{\rm for}~
k\geq 1 &(6.39) \cr }$$
It is then convenient to write the Laplace transforms in the sum rule (6.22)
in the notation
$$
{\cal L}({\cal F;G})={\cal L}({{\cal F}_{gg};{\cal G}_{gg}})+ 2\Bigl({\bar
m_s \over n_f}\Bigr){\cal L}({{\cal F}_{qg};{\cal G}_{qg}})+ \Bigl({\bar
m_s \over n_f}\Bigr)^2{\cal L}({{\cal F}_{qq};{\cal G}_{qq}}) \eqno(6.40)
$$
where the indices $gg,~qg$ and $qq$ correspond respectively to the
gluon-gluon, quark-gluon and quark-quark correlators in eq.(6.24).

\vskip0.7cm
\noindent{\bf 6.1 ~~The sum rule for $\chi'(0)$ in the chiral limit}

\vskip0.5cm

In the chiral limit, $\psi'_5(0)$ reduces to the purely gluonic correlator
$\chi'(0)$ evaluated for zero quark mass. This estimate has already been
done in our earlier paper [NSV2]. In this case, the chiral Ward identities
require $\psi_5(0) = 0$. The Laplace transforms of the gluonic correlator
read [12]: 
$$\eqalignno{
{\cal L}({{\cal F}_{gg}})&=\Bigl({\asb \over 8\pi}\Bigr)^2 {2 \over
\pi^2}\t \bigl( 1-e^{-t_c\t} (1+t_c\t)\bigr) \biggl[1+\Bigl({\asb \over
4\pi}\Bigr)\Bigl[83+4\b_1(1-\c_E) + 24 (3m_s^2 + {\pi\over\asb}\l^2)\t
\Bigr]\biggr] \cr &+\Bigl({\asb \over 8\pi}\Bigr)\t^3
\biggl[{1 \over 2\pi}\langle\as G^2\rangle + \aspb\t \langle g
G^3\rangle\biggr] &(6.41) \cr
}$$
and
$$\eqalignno{
{\cal L}({{\cal G}_{gg}})&=\Bigl({\asb \over 8\pi}\Bigr)^2 {2 \over
\pi^2}\t \bigl( 1-e^{-t_c\t}\bigr) \biggl[1+\Bigl({\asb \over
4\pi}\Bigr)\Bigl[83-4\b_1\c_E -24\c_E(3m_s^2+{\pi\over\asb}\l^2)\t
\Bigr]\biggr] \cr &-\t^3\Bigl({\asb \over 8\pi}\Bigr)
\biggl[{1 \over 2\pi}\langle\as G^2\rangle +\Bigl({\asb \over 2\pi}\Bigr)
\t \langle g G^3\rangle\biggr] &(6.42) \cr }$$
Substituting into the Laplace sum rule (6.22), we find the result shown in
Fig.~1. This gives a plot of $\sqrt{\chi'(0)}$ versus $\tau$ for the
optimal value of $t_c=6 ~\GV^2$. We observe good stability in the range of
$\tau$ from 0.2 to 0.5 ~$\GV^{-2}$.
At the stability points, we find 
$$
\sqrt{\psi'_5(0)}\big|_{m_s=0} ~\equiv~ \sqrt{\chi'(0)} ~=~ 
(26.4 \pm 4.1) ~{\MV}
\eqno(6.43)
$$
The central value is a little higher than the Laplace sum rule value of
$(22.3\pm 4.8)~{\MV}$ obtained in ref.[12]
due to the change[39] in the $O(\alpha_s^3)$ perturbative coefficient
in eq.(6.26). The different sources of errors are summarised in Table 1.

\vskip0.2cm
Since the validity of the spectral sum rules for calculating $\chi'(0)$ has
been criticised in the literature by Ioffe[16,17] (see refs.[16-20]
and our detailed rebuttal in Appendix D), we should emphasise some features of 
this derivation.

First, one should notice that the optimization of the sum rule is obtained
at the scale $\t^{-1}= (2-5) ~\GV^2$, which is relatively high compared to
the scale of ordinary mesons of the order of $m^2_\rho\simeq 0.6 ~\GV^2$.
This result is in agreement with the expectation [41,40] that the scale
of the $U(1)$ channel (gluonium) is relatively high compared with the
flavour non-singlet (meson) scale. At the optimisation scale we therefore
expect (and find) that higher dimension condensates (including `instanton'
effects[18,52]) are strongly suppressed. This is contrary to the claims in
ref.[19], where $\t^{-1}$ is taken at the too low value of $1 ~\GV^2$.

Second, the apparently large perturbative radiative corrections in the
expressions for the two-point correlators tend to cancel in the sum rule
(6.22), explaining the almost equal value of $\chi'(0)$ obtained at leading
order and the one including the perturbative $\a_s$ corrections. This is
reassuring in view of the unknown higher order radiative corrections. In
order to study the convergence of the perturbative series, we have
estimated the $\a_s^2$ corrections \`{a} la BNP [30] assuming that the
coefficient of $\as$ grows geometrically, which, numerically, is about 430.
This effect remains a small correction to the lowest order estimate.

We have also studied the effect of a $1/k^2$ correction due to the
summation of the perturbative series, which we have parametrised here (see
ref.[40]) through a phenomenological tachyonic gluon mass $\lambda^2$. We
see that the presence of this term tends to shift the optimisation scale to
smaller $\t$ values, but affects only slightly the value of $\chi'(0)$.

\vskip0.2cm
The decay constant $f_{\eta'}$ defined in eq.(6.21) can be estimated in the
same way using just the first (once-subtracted) sum rule in eq.(6.18). We
find the result (see Fig.~2) 
$$
f_{\eta'} = (24.4 \pm 3.6) ~\MV
\eqno(6.44)
$$
where the different sources of error are again given in Table 1.

\vskip0.2cm

One can also derive a Finite Energy Sum Rule (FESR) for
$\chi'(0)$. This can simply be found by taking the small $\t$
limit of the Laplace transform sum rule. Including the radiative
corrections, and neglecting the small $\l^2$ corrections, we obtain [12]
$$
\chi'(0)~\simeq ~ \int_0^{t_c} {dt\over t^2} {1\over\pi} {\rm Im}\psi_5(t) 
- \Bigl({\asb\over 8\pi}\Bigr)^2 {2\over \pi^2} t_c
\biggl[ 1+ \Bigl({\asb \over
4\pi}\Bigr)\bigl(83-4\b_1\bigr)\biggr] \eqno(6.45)
$$
which confirms the consistency of the set of values of the parameters
$\chi'(0), ~f_{\eta'}$ and $t_c$ obtained from the Laplace sum rules.

\vskip0.7cm
\noindent{\bf 6.2 ~~The sum rule for $\psi'_5(0)$ with massive quarks}

\vskip0.5cm
In the case of massive quarks, $\psi_5(0)$ is non-zero. However, its value
is known from the exact chiral Ward identities (2.9) or (2.11): 
$$
\psi_5(0) ~=~ - {4\over(2n_f)^2} m_s \langle\bar ss\rangle \eqno(6.46)
$$
where we again set $m_u = m_d =0$. As before, we then have two unknown
physical quantities, viz. ${\rm Im}\psi_5(t)$ and $\psi'_5(0)$, to be
determined from the two subtracted sum rules (6.18) and (6.19). (The
unsubtracted sum rule[29], independent of $\psi_5(0)$, is more sensitive
to the higher meson and gluonium mass than the subtracted sum rules, but is
in any case not needed in this analysis.)

\vskip0.3cm
We now need the Laplace transforms of the quark-gluon and quark-quark
correlators (see eqs.(6.31) and (6.32)). First, for the quark-gluon
correlator, we have

$$\eqalignno{
{\cal L}({{\cal F}_{qg}})~=~&\t^2\biggl[ \aspb^2{3m_s \over
16\pi^2}\Bigl({11 \over 6}\Bigr) +\c_E{\t \over 2}\aspb^2\langle\bar
ss\rangle \cr &+\aspb {m_s \over 8\pi} \langle\as G^2\rangle(1-\c_E)\t^2
\biggr]\bigl( 1-e^{-t_c\t}\bigr)
+ \aspb {\t^2\over2} M_0^2 \langle\bar s s\rangle &(6.47) \cr }
$$
and
$$\eqalignno{
{\cal L}({{\cal G}_{qg}})~=~&\t^2\biggl[ \aspb^2{3m_s \over
16\pi^2}\Bigl({11 \over 6}-{17 \over 3}\c_E+
3\c_E^2+\zeta(2)\Bigr)+(1-\c_E){\t \over 2}\aspb^2\langle\bar ss\rangle \cr
&+\aspb {m_s \over 8\pi} \langle\as G^2\rangle\Bigl(\c_E-{3 \over 2}\Bigr)
{\t^2 \over 2}\biggr] \bigl( 1-e^{-t_c\t}\bigr) - \aspb {\t^2\over4} M_0^2
\langle\bar ss\rangle &(6.48) \cr
}$$

The Laplace transform of the perturbative part of the quark-quark
correlator including the $O(\as)$ corrections has a non-trivial analytic
expression in terms of the Riemann $\zeta$ functions. It is more convenient
to express it in a numerical form: 
$$\eqalignno{
{\cal L}({\cal F}_{qq}^{PT})&={3 \over 8\pi^2}\t^2\biggl[ 1+\Bigl({17 \over
3}+2\c_E\Bigr)\aspb+O(\alpha_s^2)\biggr] \bigl(1-e^{-t_c\t}\bigr) \cr
{\cal L}({\cal G}_{qq}^{PT})&=-{3 \over 8\pi^2}\t^2\biggl[
2+\c_E+13.298\aspb+O(\alpha_s^2)\biggr] \bigl( 1-e^{-t_c\t}\bigr) &(6.49)
\cr }$$
The Laplace transforms of the $m^2_s$ and $\l^2$ corrections are 
$$\eqalignno{
{\cal L}({\cal F}^{(2)}_{qq})&={3 \over 8\pi^2}\t^3\biggl[
\bm^2\biggl[2\c_E-12.868\aspb\biggr] - 4\c_E \aspb \l^2 \biggr] \cr {\cal
L}({\cal G}^{(2)}_{qq})&={3 \over 8\pi^2}\t^3\biggl[
\bm^2\biggl[2(1-\c_E)+12.152\aspb\biggr] +4(\c_E - 1)\aspb \l^2 \biggr]
&(6.50) \cr
}$$
and the Laplace transforms of its non-perturbative part are 
$$\eqalignno{
{\cal L}({\cal F}_{qq}^{NP})~=~&-\t^4\biggl[ {3 \over
8\pi^2}\bm^4(1+2\c_E)+\Bigl({1 \over 8\pi} \langle\as G^2\rangle-m_s\langle
\bar ss\rangle\Bigr) \cr &+{\t \over 2}\Bigl[ -M^2_0 m_s\langle\bar
ss\rangle+ {112 \over 27}\pi\rho\as\langle\bar ss\rangle^2\Bigr]\biggr] \cr
{\cal L}({\cal G}_{qq}^{NP})~=~&\t^4\biggl[ {3 \over 8\pi^2}\bm^4\c_E+ {1
\over 2}\Bigl({1 \over 8\pi}\langle \as G^2\rangle - m_s\langle\bar
ss\rangle\Bigr) \cr
&-{\t \over 6}\Bigl[ -M^2_0 m_s\langle\bar ss\rangle+ {112 \over
27}\pi\rho\as\langle\bar ss\rangle^2\Bigr]\biggr] &(6.51) \cr }
$$

\vskip0.3cm
Finally, collecting all these expressions in the combination (6.40), we
deduce the Laplace sum rule for the complete correlation function
$\psi'_5(0)$ with $m_s\neq0$. Our result is shown in Fig.~1. We find
$$
\sqrt{\psi'_5(0)} = (33.5 \pm 3.9) ~\MV
\eqno(6.52)
$$
for the range of stability in $\tau = (0.2 - 0.4) ~\GV^{-2}$. The result
obtained with $\l^2 = 0$ is very similar, though the stability in $\tau$
occurs at a slightly larger value of $\t \sim 0.6 ~\GV^{-2}$.

Similarly, we find the value for the decay constant $f_{\eta'}$ for
non-zero strange quark mass (see Fig.~2): $$
f_{\eta'} = (27.4 \pm 3.7) ~\MV
\eqno(6.53)
$$

Comparing these results with those obtained above in the chiral limit,
we find that the effect of the $SU(3)$ breaking quark mass is to increase
the values of $f_{\eta'}$ and $\sqrt{\psi'_5(0)}$ by approx.~10\% and 20\%
respectively.
This is a reasonable conclusion. The $SU(3)$ breaking effects are not
negligible, but they are of the order expected in relatively smooth
quantities such as decay constants which, as explained in section 5, are
expected to be only weakly dependent on the quark masses.

Certainly we find no evidence of the huge $SU(3)$ breakings advertised in
ref.[19], which were taken as an indication of the failure of the spectral
sum rule method in the flavour singlet channel. The main reason for our
different conclusion is that the $\t$ stability region in our calculation
is found to be much lower than that used in ref.[19], and so the $SU(3)$
breaking terms of $O(m_s^2\t)$ are much smaller implying a much better
convergence of the corresponding OPE. A more detailed comparision of our
work with ref.[19] is given in appendix D.

\vskip0.7cm
\noindent{\bf 6.3 ~~Flavour non-singlet correlation functions}

\vskip0.5cm

For the full unified Goldberger-Treiman relations, we also need the
corresponding results for the octet current. These results are immediately 
obtained from the formulae presented above,
with the obvious changes of $m_{\eta}$ for $m_{\eta'}$ etc.
In the derivation, we find it convenient to use the normalisations:
$$
\langle 0|\pl^\m J_{\m 5}^8|\eta\rangle = 
{{2n_f}\over{\sqrt{3}}} ~\sqrt2 ~\tilde f_{\eta} m_{\eta}^2,
\eqno(6.54)
$$
$$
\psi_5^{88}(k^2) = {{3}\over{(2n_f)^2}} ~i\int
d^4x~e^{ik.x}~\langle0|T~D^8(x)~D^8(0)|0\rangle
\eqno(6.55)
$$
and:
$$
\psi_5^{08}(k^2) = {{\sqrt 3}\over{2n_f}} ~i\int
d^4x~e^{ik.x}~\langle0|T~D^0(x)~D^8(0)|0\rangle
\eqno(6.56)
$$
where $D^a$ is normalised as in eq.(2.10). 
With these normalisation factors, the quark correlator $\psi_{ss}$ 
occurs with the same coefficient in the sum rules for 
$\psi_5$, $\psi_5^{88}$ and $\psi_5^{08}$. 
The octet decay constant is related to the conventionally normalised 
$f_{\eta}$ by
$$
f_{\eta} = 2\sqrt6~\tilde f_{\eta}
\eqno(6.57)
$$
so that $f_{\eta}=f_\pi=93.3$ MeV for exact SU(3).
We find:
$$
\sqrt{\psi_5^{\prime 88}(0)} = (43.8 \pm 5.0) ~\MV 
\eqno(6.58)
$$
and
$$
\tilde f_{\eta} = (30.0 \pm 3.4) ~\MV 
\eqno(6.59)
$$
which corresponds to 
$$
f_{\eta} = (147 \pm 17) ~\MV 
\eqno(6.60)
$$
The sources of the errors are again tabulated in Table 1. 

In order to quantify the systematic errors of the approach, we have
re-estimated the value of $f_\pi$ using the same inputs, approximations and
methods (for different estimates of $f_\pi$ from sum rules, see e.g.~[29]) 
as used above for $f_\eta$, by using obvious changes of the parameters 
(quark masses, meson mass, continuum threshold). In this way, one obtains:
$$
f_{\pi} = (107 \pm 12) ~\MV 
\eqno(6.61)
$$
Taking into account the slight deviation of the central value from the
experimental number, we can consider as a final result:
$$
f_{\eta}/f_{\pi} = 1.37 \pm 0.16 
\eqno(6.62)
$$
where we expect that the error quoted here has been over-estimated.
This ratio is in line with our expectations, since phenomenologically
$f_K \simeq 1.2 f_\pi$, and we would expect $SU(3)$ breaking to be stronger 
for the $\eta$ than the $K$ [23,24].

As a by-product, we have checked that a systematic rescaling of the value 
of the decay constants $f_\eta$ and $f_{\eta'}$ will affect similarly the 
value of the slope of both the non-singlet $\psi_5^{\prime 88}(0)$ and singlet 
$\psi'_5(0)$ correlators, such that the ratios of correlators and decay 
constants which we use are not affected by this change.

Together with good $t_c$ and $\t$ stability, these results therefore
confirm the general reliability of the spectral sum rules in this channel. 
\vskip0.4cm

\centerline{\eightpoint
\vbox{\offinterlineskip
\hrule
\halign{&\vrule#&
\strut\quad\hfil#\quad\cr
height4pt&\omit&&\omit&&\omit&&\omit&&\omit&&\omit&&\omit&\cr
&\hfil $\D O[\MV]$ && \hfil $t_c=6\pm2$ && \hfil $\L$ && \hfil $\bar{m}_s$
&& \hfil
$\t=0.3\pm0.1$ && \hfil $\langle \a_s G^2\rangle$ && \hfil Total &\cr
height4pt&\omit&&\omit&&\omit&&\omit&&\omit&&\omit&&\omit&\cr \noalign{\hrule}
height6pt&\omit&&\omit&&\omit&&\omit&&\omit&&\omit&&\omit&\cr
&$f_{\eta'}|_{m_s=0}$ && 2.5 &&
2.3 && -- && 0.9 && 0.8 && 3.6 &\cr
height6pt&\omit&&\omit&&\omit&&\omit&&\omit&&\omit&&\omit&\cr
&$\sqrt{\chi'(0)}$ && 3.3 && 2.2
&& -- && 0.1 && 0.8 && 4.1 &\cr
height6pt&\omit&&\omit&&\omit&&\omit&&\omit&&\omit&&\omit&\cr
&$f_{\eta'}$ && 2.7 && 2 && 1.3 && 0.5 && 0.8 && 3.7 &\cr
height6pt&\omit&&\omit&&\omit&&\omit&&\omit&&\omit&&\omit&\cr
&$\sqrt{\psi_5^{\prime}(0)}$ &&
2.1 && 1.9 && 2.6 && -- && 0.8 && 3.9 &\cr
height6pt&\omit&&\omit&&\omit&&\omit&&\omit&&\omit&&\omit&\cr &$\tilde
f_{\eta}$ && 0.9 && 1.3
&& 2.9 && -- && 0.8 && 3.4 &\cr
height6pt&\omit&&\omit&&\omit&&\omit&&\omit&&\omit&&\omit&\cr
&$\sqrt{\psi_5^{\prime 88}(0)}$&& 1.3 && 2 && 4.3 && 0.3 && 0.8 && 5.0 &\cr
height6pt&\omit&&\omit&&\omit&&\omit&&\omit&&\omit&&\omit&\cr
&$f_{\eta'}/{\tilde f_{\eta}}$ &&
0.030 && 0.025 && 0.045 && -- && -- && 0.060 &\cr
height6pt&\omit&&\omit&&\omit&&\omit&&\omit&&\omit&&\omit&\cr
&${\sqrt{\psi_5^{\prime}(0)}\over{\sqrt{\psi_5^{\prime 88}(0)}}}$ && 0.022
&& 0.008 && 0.015 &&
-- && -- && 0.028 &\cr
height8pt&\omit&&\omit&&\omit&&\omit&&\omit&&\omit&&\omit&\cr } \hrule} }

\vskip0.2cm
\centerline{\eightpoint Table 1: ~~Error estimates in $\MV$ for the
different observables.}
\centerline{\eightpoint The sources of errors are in $[\GV]^d$ where $d$ is
the corresponding dimension.}

\vskip0.3cm

As already mentioned in sections 3 and 4,
the calculation of the off-diagonal correlator $\psi_5^{\prime 08}$ 
is much more delicate, since the contributions of both the $\eta$ and $\eta'$
to the spectral function must be taken into account. These tend to 
cancel because of the relative signs of the decay constants after
flavour mixing, and the sum rule prediction for $\psi_5^{\prime 08}$
should then be relatively small. This is confirmed by preliminary estimates.
However, a complete calculation including the effects of flavour mixing
for both the off-diagonal $\psi_5^{\prime 08}$ and diagonal
$\psi_5^{\prime}$ and $\psi_5^{\prime 88}$ correlators requires
significant further analysis and is beyond the scope of the present work.

\vskip0.2cm
We conclude that the QCD spectral sum rule method is indeed giving reliable
results for the decay constants and susceptibilities in both the flavour
singlet and non-singlet channels. Although the pattern of cancellations of
quark mass effects observed in the effective lagrangian, or pole dominance,
analysis in section 5 is not manifest in the more precise sum rule method,
the essential observation that the slope at $k=0$ of the correlation
functions $W_{S_D^a S_D^b}$ is relatively insensitive to the quark masses 
is confirmed by the numerical results found here.

\vfill\eject

\noindent{\bf 7. ~~Quantitative Analysis of the Unified GT Relation} 
\vskip0.5cm
In this section, we discuss the implications of these 
numerical results for the unified Goldberger-Treiman relations and 
the `proton spin' suppression.

Collecting the results of the previous section and re-converting
to the normalisations in sections 2-5, we have found
$$
\bigl(FF^T\bigr)_{ab} ~=~ \lim_{k=0} ~{d\over dk^2} ~W_{S_D^a S_D^b} ~=~
\left(\matrix{(201 ~\pm~ 23)^2 & {} \cr {} & 
(152 ~\pm~ 17)^2 \cr}\right) ~\MV^2
\eqno(7.1)
$$
in the $a,b = 0,8$ sector, within an approximation where we have
kept only the $\eta'$, or $\eta$, in the spectral functions 
for the correlators $\langle 0|T~D^0 ~D^0|0\rangle$, or 
$\langle 0|T~D^8~D^8|0\rangle$, respectively.

Within this approximation, neglecting $SU(3)$ flavour mixing
in the unified GT formula, the generalisation beyond the chiral limit 
of the suppression formula for the singlet axial charge is (see eq.(4.10)):
$$
{a^0 \over a^8}  ~~=~~
{1\over \sqrt6}~ {F_{00} \over F_{88}} ~~=~~
{1\over \sqrt2} {\sqrt{\psi'_5(0)} \over \sqrt{\psi_5^{\prime 88}(0)}}~~=~~ 
0.55 ~\pm~ 0.02
\eqno(7.2)
$$
This should be compared with the corresponding result in the chiral
limit, using our new determination (6.43) of $\chi'(0)$ and using the sum 
rule estimate (6.61) for $f_{\pi}$:
$$
{a^0 \over a^8}  ~~=~~
{\sqrt6\over f_\pi}~\sqrt{\chi'(0)} ~~=~~ 0.60 ~\pm~ 0.12
\eqno(7.3)
$$
The relatively small error in eq.(7.2) is due to the cancellation of 
the systematic errors in the ratio (see Table 1), which was not taken 
into account in eq.(7.3).
Running these results from the scale $\t^{-1} \simeq 3 ~\GV^2$ 
to the SMC scale of $Q^2 = 10 ~\GV^2$, and substituting
$a^8 = 3F-D = 0.58 \pm 0.03$, we find
$$\eqalignno{
a^0(Q^2=10\GV^2) &= 0.31 \pm 0.02 \cr
\C_1^p(Q^2=10\GV^2) &= 0.141 \pm 0.005
&(7.4) \cr }
$$
compared with
$$\eqalignno{
a^0(Q^2=10\GV^2) &= 0.33 \pm 0.05 \cr
\C_1^p(Q^2=10\GV^2) &= 0.144 \pm 0.009
&(7.5) \cr }
$$
in the chiral limit.

We therefore find very good agreement between the final prediction
for the singlet axial charge in the presence of quark masses and in the
chiral limit. This confirms our theoretical expectation that 
$a^0$ is relatively insensitive to the quark masses.
Moreover, our new prediction for the `proton spin' suppression remains,
notwithstanding the large errors on the experimental data, in good
agreement with the experimental results quoted in section 4.

\vfill\eject

\noindent{\bf 8. ~~Conclusions} 
\vskip0.3cm
In this paper, we have extended our previous analysis of the `proton spin'
problem in the chiral limit by considering the effects of chiral
$SU(3)$ symmetry breaking and flavour mixing due to the quark masses.

The formal basis of our analysis is the derivation of the new, unified
Goldberger-Treiman relations:
$$
G_A^a ~=~  {1\over 2m_N}~F_{ab} ~\hat \C_{\eta^b NN} 
\eqno(8.1)
$$
where $F$ is determined from 
$$
F_{ac} F_{cb}^T = \lim_{k=0} {d\over dk^2}~ i \int dx~e^{ikx}~\langle 0|
T~\pl^\m J_{\m5}^a(x)~\pl^\n J_{\n5}^b(0)|0\rangle
\eqno(8.2)
$$
Apart from the mixing between the flavour
singlet and octet channels induced by the non-vanishing strange quark mass,
the most significant change is the generalisation of the slope of the
topological susceptibility $\chi'(0)$ to the equivalent correlation 
function involving the {\it total} divergence of the singlet axial
current, viz. ${d\over dk^2} \langle0|T~\pl^\m J_{\m 5}^0 ~
\pl^\n J_{\n 5}^0 |0\rangle\big|_{k=0}$. This is the quantity which displays
the smoothest approach to the chiral limit, with the explicit quark
mass dependence present in the individual correlators (section 5)
cancelling in the sum. This observation may have important implications
for attempts to calculate the topological susceptibility using 
lattice methods[53], where it is notoriously difficult to approach the
chiral limit too closely.

We emphasise again that the unified GT relations are new results and
are {\it exact} in QCD. The familiar PCAC forms (in the flavour
non-singlet channels) are obtained by approximating the 1PI vertices
by the corresponding low-energy meson-nucleon coupling constants
and by approximating the slopes of the current correlation functions
by decay constants. Away from the chiral limit, both approximations
assume pole-dominance of the matrix elements and correlation functions
by the pseudo-Goldstone bosons.

The relevant correlation functions were then evaluated using QCD spectral 
sum rules. In the singlet as well as octet channel, very good stability
was obtained with respect to the parameter $\t$. The optimum
value of $\t^{-1} = 2.5-5~\GV^2$ is in line with general arguments
for the appropriate scale in the gluon-rich $U_A(1)$ channel.
Specific criticisms[16-20] of the applicability of spectral sum rules 
to the $U_A(1)$ channel were shown to be incorrect.

Our final numerical results agree with our earlier findings[12].
The suppression in the flavour singlet axial charge, driven by the 
mechanism of topological charge screening, is confirmed. The reduction
in the predicted value of $a^0(Q^2)$ compared to its chiral limit
is slight (less than $10\%$), in agreement with our general theory,
although it should be emphasised that our results still neglect
flavour mixing.
Translated into a prediction for the first moment of  
$g_1^p$, under the assumption that the
RG-invariant vertices in eq.(8.1) are well approximated by their OZI
values, our final result:
$$\eqalignno{
a^0(Q^2=10\GV^2) &= 0.31 \pm 0.02 \cr
\C_1^p(Q^2=10\GV^2) &= 0.141 \pm 0.005
&(8.3) \cr }
$$
remains in good agreement with experiment, and confirms our proposal that
the `proton spin' suppression is a target-independent effect due to the
screening of topological charge by the QCD vacuum.

\vfill\eject

\noindent{\bf Appendix A:~~Partial Legendre transforms} 
\vskip0.5cm
We derive here the relations between the Green functions and vertices,
defined as functional derivatives of the generating functionals $W$ and
$\C$ respectively, used in the derivation of the GT relations[11].

For generality, we derive these results for a set of fields $\Phi^a$ and
`currents' $J^r$, with sources $S^a$ and $V^r$, where the partial Legendre
transform is made wrt the sources $S^a$ only. That is, 
$$
W[V,S] = \C[V,\Phi] + S^a \Phi^a
\eqno(A.1)
$$
We adopt a compact notation where any Lorentz indices are implicit and a
spacetime integration is assumed in the sum over repeated indices. As in
the text, functional differentiation is indicated by subscripts. Thus,
e.g., $W_S$ denotes ${\d W\over \d S}$ at fixed $V$, etc.

By definition,
$$
\Phi^a = W_{S^a}
\eqno(A.2)
$$
while
$$
\C_{\Phi^a} = - S^a
~~~~~~~~~~~~~~~~~
\C_{V^r} = W_{V^r}
\eqno(A.3)
$$
From
$$\eqalignno{
\d^a_b ~=~ {\d\over \d \Phi^a} \Phi^b ~&=~ {\d V^r\over \d\Phi^a}~ W_{V^r
S^b} ~+~ {\d S^c\over \d\Phi^a}~ W_{S^c S^b} \cr &{}\cr
&=~ - \C_{\Phi^a \Phi^c}~ W_{S^c S^b}
&(A.4) \cr }
$$
we recover the usual result that the 2-point vertex matrix is just the
inverse of the propagator matrix, but in this case restricted to the $S^a,
\Phi^a$ sector.

Similarly, from
$$\eqalignno{
0 ~=~ {\d\over\d V^r} \Phi^b ~&= ~{\d V^t\over\d V^r}~ W_{V^t S^b} ~+~ {\d
S^c\over\d V^r}~ W_{S^c S^b} \cr
&{}\cr
&=~ W_{V^r S^b}~ - ~\C_{V^r \Phi^c}~ W_{S^c S^b} &(A.5) \cr }
$$
we find
$$
\C_{V^r \Phi^b} = W_{V^r S^c}~ W_{S^c S^b}^{-1} \eqno(A.6)
$$
which can therefore be identified as the matrix element $\langle
0|J^r|\Phi^b\rangle$.

Taking this further,
$$\eqalignno{
0 ~&=~ {\d\over\d V^r} \Bigl(W_{S^a S^c} \C_{\Phi^c \Phi^b} \Bigr) \cr &{}\cr
&=~ W_{V^r S^a S^c}~ \C_{\Phi^c \Phi^b} + W_{S^a S^c}~ \Bigl( \C_{V^r
\Phi^c \Phi^b} ~+~ W_{V^r S^d}~ \C_{\Phi^d \Phi^c \Phi^b} \Bigr) &(A.7) \cr
}
$$
from which we find
$$
\C_{V^r \Phi^c \Phi^b} ~+~ W_{V^r S^d}~ \C_{\Phi^d \Phi^c \Phi^b}~ =~
W_{S^a S^c}^{-1}~ W_{V^r S^c S^d}~ W_{S^d S^b}^{-1} \eqno(A.8)
$$
which is identified as $\langle \Phi^a|J^r|\Phi^b\rangle$.

As a final example, we derive the crucial identity (3.11) used in our
derivation of the GT relations:
$$\eqalignno{
\C_{V^r V^s} ~=~ {\d\over\d V^r} W_{V^s} ~&=~ {\d V^t\over\d V^r} W_{V^t
V^s} ~+~ {\d S^d\over\d V^r} W_{S^d V^s} \cr &{}\cr
&=~ W_{V^r V^s}~ +~ \C_{V^r \Phi^d} W_{S^d V^s} \cr &{}\cr
&=~ W_{V^r V^s}~ -~ W_{V^r S^c}~ W_{S^c S^d}^{-1}~ W_{S^d V^s} &(A.9) \cr }
$$
the last line following from eq.(A.6).

\vskip1cm

\noindent{\bf Appendix B:~~Current algebra, Dashen's formula and the GT
relation}
\vskip0.5cm
As an illustration of how standard current algebra (PCAC) relations arise
from our formalism, we give here a short derivation of the Dashen formula
for the masses of the pseudo-Goldstone bosons.

The starting point is the identification (see eq.(A.6)) $$
\C_{V_{\m5}^a \eta^b} = \langle 0|J_{\m5}^a|\eta^b\rangle = ik_\m
f^{ab}(k^2) \eqno(B.1)
$$
where the on-shell function $f^{ab}(m_{\eta}^2)$ is the decay constant
matrix for the pseudo-Goldstone bosons $\eta^a$. In terms of the normalised
fields $\eta^a = B_{ab}\phi_5^b$, the Ward identity (2.14) is written as $$
ik_\m \C_{V_{\m5}^a \eta^b} + \Phi_{ac} B_{cd}^T ~\C_{\eta^d\eta^b} -
M_{ac} B_{cb}^{-1} = 0
\eqno(B.2)
$$

Now take ${d\over dk^2}\big|_{k=0}$ of this equation. Using the
normalisation condition (3.7), we find immediately $$
f_{ab}(0) = \Phi_{ac} B_{cb}^T
\eqno(B.3)
$$
The assumption that $f_{ab}$ is only a slowly-varying function of $k^2$
then allows (B.3) to be identified with the decay constant matrix. This is
(as explained in detail in ref.[11]) the standard PCAC approximation,
equivalent to pole dominance of correlation functions by pseudo-Goldstone
bosons.

In the same approximation, we can write
$$
\C_{\eta^a \eta^b} = k^2 \d_{ab} - (m_\eta^2)_{ab} \eqno(B.4)
$$
Neglecting the mixing with Q, this would produce a pole in the propagator
matrix with the corresponding mass matrix obtained from the zero-momentum
limit of the Ward identity (B.2). In matrix notation, this means that
$$
0 = - \Phi B^T m_\eta^2 - M B^{-1}
\eqno(B.5)
$$
and so
$$
\Phi B^T m_\eta^2 B \Phi^T = - M B^{-1} B \Phi^T \eqno(B.6)
$$
Using the identification (B.3) of the decay constant, we therefore have
(recall $\Phi^T = \Phi$),
$$
f_{ac}~ (m_\eta^2)_{cd} ~f_{db}^T = - M_{ac} \Phi_{cb} \eqno(B.7)
$$
This is Dashen's formula for the pseudo-Goldstone bosons, which for the
$n_f=2$ chiral symmetry breaking
pattern $SU(2)_L \times SU(2)_R \rangle SU(2)_V$ is simply $$
f_\pi^2~ m_\pi^2 = - (m_u + m_d) \langle \bar q q\rangle \eqno(B.8)
$$

Mixing with the glueball field $Q$ in general shifts the masses of the
physical pseudo-Goldstone bosons. However, in the simple case where
$\Phi_{a0} \equiv 2\langle\phi^a\rangle$ is zero for $a \neq 0$, this is
only relevant in the singlet sector. There, the formula (B.7) refers to the
unphysical $\eta^0$, {\it not} to the $\eta'$. The corresponding formula
for the physical $\eta'$ including mixing with the glueball field $Q$, and
involving a different identification of the $\eta'$ decay constant, is
derived (in the chiral limit) in ref.[11].

We can also relate the generalised GT relations in the text to the standard
formula using pole dominance. For $n_f=2$, eq.(3.18) reduces to $$
m_N ~g_A =  F~ g_{\pi NN}
\eqno(B.9)
$$
where $g_A \equiv 2 G_A^3$ in our notation. Evaluating $F$ from eq.(3.20)
using pole dominance, we find
$$
F^2 = {d\over dk^2} \Bigl(\big|\langle 0|\pl^\m J_{\m5}^3|\pi\rangle\big|^2
{(-1)\over k^2-m_\pi^2} \Bigr)\Big|_{k=0} = f_\pi^2 \eqno(B.10)
$$
recovering the standard GT formula and confirming the interpretation of $F$
in eq.(3.18) as a decay constant matrix. Notice that since this derivation
assumes pole dominance, it is an approximation. Unlike the new relation
(3.18), the standard GT relation becomes exact only in the chiral limit.

\vskip1cm

\noindent{\bf Appendix C:~~ Renormalisation Group} 
\vskip0.5cm

The renormalisation group equations (RGEs) for the various quantities
arising in the derivation of the unified GT relations can be derived using 
the same methods developed in ref.[11]. In this appendix, we summarise
the most important identities.

The starting point is the definition of the renormalised composite
operators[54]. For QCD, with non-zero quark masses, we have
(denoting bare operators with the label `$B$' and renormalised operators
with no label for simplicity)
$$\eqalignno{
&J_{\m5}^0 = Z J_{\m5}^{0B}  ~~~~~~~~~~
J_{\m5}^{a\neq0} =  J_{\m5}^{a\neq0 B} \cr
&Q = Q^B - {1\over 2n_f}(1-Z) \pl^\m J_{\m 5}^{0B} \cr
&\phi_5^a = Z_\phi \phi_5^{aB} ~~~~~~~~~~~~~
\phi^a = Z_\phi \phi^{aB} 
&(C.1) \cr}
$$
where $Z_\phi$ is the inverse of the mass renormalisation, 
$Z_\phi = Z_m^{-1}$. The anomalous dimensions associated with $Z$ and 
$Z_\phi$ are denoted $\c$ and $\c_\phi$ respectively. These definitions
ensure that the combinations $\pl^\m J_{\m5}^0 - 2n_f Q$ and
$m_q [\bar q \c_5 q]$ occurring in the $U_A(1)$ anomaly equation
(e.g.~eq.(6.2)) are RG invariant.

The fundamental RGE for the generating functional $W$ is therefore
(in the notation of section 2, where suffices on $W$ denote 
functional differentiation):
$$
\DD W = \c\Bigl(V_{\m5}^0 - {1\over2n_f}\pl_\m \theta\Bigr)W_{V_{\m5}^0}
+ \c_\phi\Bigl(S_5^a W_{S_5^a} + S^a W_{S^a}\Bigr) + \ldots
\eqno(C.2)
$$
where $\DD = \Bigl(\m{\pl\over\pl\m} + \b{\pl\over\pl g} - \c_m\sum_q
m_q{\pl\over\pl m_q}\Bigr)\Big|_{V,\theta,S_5,S}$ and a spacetime
integration is assumed in the sum over repeated indices.
The notation $+\ldots$ refers to the additional terms which are required 
to produce the contact term contributions to the RGEs for $n$-point
Green functions of composite operators. These are discussed fully
in ref.[11], but will be omitted here for simplicity. They vanish at 
zero momentum.

The RGEs for Green functions are found simply by differentiating
eq.(C.2) wrt the sources. Simplifying the results using the
chiral Ward identities (2.5), we find a complete set of RGEs for the
2-point functions. These are:
$$\eqalignno{
&\DD W_{V_{\m5}^0 V_{\n5}^0} = 2\c W_{V_{\m5}^0 V_{\n5}^0} +\ldots ~~~~~
\DD W_{V_{\m5}^0 V_{\n5}^b} = \c W_{V_{\m5}^0 V_{\n5}^b} +\ldots~~~~~
\DD W_{V_{\m5}^a V_{\n5}^b} = 0 +\ldots \cr
&\DD W_{V_{\m5}^0 \theta} = 2\c W_{V_{\m5}^0 \theta} 
+ \c {1\over2n_f} M_{0b} W_{V_{\mu5}^0 S_5^b} +\ldots \cr
&\DD W_{V_{\m5}^a \theta} = \c W_{V_{\m5}^a \theta}
+ \c {1\over2n_f} M_{0b} W_{V_{\mu5}^0 S_5^b} +\ldots    \cr
&\DD W_{V_{\mu5}^0 S_5^b} = (\c + \c_\phi) 
W_{V_{\mu5}^0 S_5^b} +\ldots ~~~~~~~~~
\DD W_{V_{\mu5}^a S_5^b} = \c_\phi W_{V_{\mu5}^a S_5^b} +\ldots \cr
&\DD W_{\theta \theta} = 2\c W_{\theta \theta} 
+ 2\c {1\over2n_f} M_{0b} W_{\theta S_5^b} +\ldots \cr
&\DD W_{\theta S_5^b} = (\c + \c_\phi) W_{\theta S_5^b} 
+ \c {1\over2n_f} \bigl(M_{0c} W_{S_5^c S_5^b}  + \Phi_{0b}\bigr) 
+\ldots \cr
&\DD W_{S_5^a S_5^b} = 2\c_\phi W_{S_5^a S_5^b} +\ldots
&(C.3) \cr}
$$
It is straightforward to check the self-consistency of these RGEs 
with the Ward identities (2.5) and (2.11). The pattern of cancellations 
which ensures this is nevertheless quite intricate.

Next, we need the RGE for the generating functional of the 1PI vertices.
This follows immediately from its definition in eq.(2.12) and the 
RGE (C.2) for $W$:
$$
\tilde\DD\C = \c\Bigl(V_{\m5}^0 - {1\over2n_f}\C_Q\pl_\m\Bigr)
\C_{V_{\m5}^0} - \c_\phi\Bigl(\phi_5^a \C_{\phi_5^a} + 
\phi^a \C_{\phi^a}\Bigr) +\ldots
\eqno(C.4)
$$
where $\tilde\DD = \Bigl(\m{\pl\over\pl\m} + \b{\pl\over\pl g} - \c_m\sum_q
m_q{\pl\over\pl m_q}\Bigr)\Big|_{V,Q,\phi_5,\phi}$.

The RGEs for the 1PI vertices are found by differentiation and, using
the Ward identities (2.14) to simplify the results, we find
in particular:
$$
\eqalignno{
&\DD \hat\C_{QNN} = - \c \hat\C_{QNN}
+ \c {1\over2n_f} \Phi_{0b} \Bigl(\C_{Q \phi_5^b} \hat\C_{QNN}
+ \C_{QQ} \hat\C_{\phi_5^b NN} \Bigr) +\ldots \cr
&\DD \hat\C_{\phi_5^a NN} = - \c_\phi \hat\C_{\phi_5^a NN}
+ \c {1\over2n_f} \Phi_{0b} \Bigl(\C_{\phi_5^a \phi_5^b} \hat\C_{QNN}
+ \C_{\phi_5^a Q} \hat\C_{\phi_5^b NN} \Bigr) 
- \c {1\over2n_f}M_{0a} \hat\C_{QNN} + \ldots \cr
&{}&(C.5) \cr }
$$
Here, $\DD = \tilde\DD + \c_\phi \langle\phi^a\rangle{\d\over\d\phi^a}$.
As explained in ref.[11], this is identical to the RG operator
$\DD$ defined above (acting on $W$) when the sources are set to 
zero and the fields to their VEVs.
With some calculation, the consistency of the GT formulae (3.4)
and (3.5) can now be shown using the RGEs (C.3) and (C.5).
Again, a very intricate pattern of cancellations occurs to ensure
this.

The RGEs for the 2-point vertices occurring in the Ward identities (2.14)
are easily found by differentiating eq.(C.4). The most important is
$$
\DD\C_{\phi_5^a\phi_5^b} = -2\c_\phi \C_{\phi_5^a\phi_5^b}
+ \c {1\over2n_f} \Bigl(\C_{\phi_5^a Q} \bigl(
\Phi_{0c}\C_{\phi_5^c\phi_5^b} - M_{0b}\bigr) + ~a\leftrightarrow b~\Bigr)
+ \ldots
\eqno(C.6)
$$

This allows us to deduce the RGE for the matrix $B_{ab}$ which relates the
$\phi_5^a$ fields to the canonically normalised $\eta^a$ fields by
$\eta^a = B_{ab}\phi_5^b$.
Recall the definition (3.9):
$$
{d\over dk^2} \C_{\phi_5^a \phi_5^b}\big|_{k=0} = B_{ac}^T ~{d\over
dk^2}\C_{\eta^c \eta^d}\big|_{k=0} ~B_{db} = B_{ac}^T B_{cb}
\eqno(C.7)
$$
From eq.(C.6) and the zero-momentum limit of the Ward identities (2.14),
we deduce
$$
\DD B_{ab} = -\c_\phi B_{ab} + \c B_{ac} \Phi_{c0} \Phi_{0b}^{-1}
\eqno(C.8)
$$

The RGE for $F_{ab}$ now follows immediately from its definition
$F = \Phi B^T$ and the RGE $\DD \Phi = \c_\phi \Phi$.
It is simply
$$
\DD F_{ab} = \c \d_{a0} F_{0b}
\eqno(C.9)
$$
that is,
$$\eqalignno{
&\DD F_{0b} = \c F_{0b} ~~~~~{\rm for~ all} ~b \cr
&\DD F_{ab} = 0 ~~~~~~~~~~a\neq 0 
&(C.10) \cr }
$$

The final step in proving RG consistency of the unified GT formulae
is to show that the vertices $\hat \C_{\eta^a NN}$ (at ${k=0}$) 
are RG invariant.
Eq.(C.10) then ensures the required RGE for the axial charges, viz.
$$
\DD G_A^a = \c \d_{a0} G_A^a
\eqno(C.11)
$$
To check this explicitly, notice that eq.(C.5) for $\hat \C_{\phi_5^a NN}$
simplifies at $k=0$. The contact terms vanish and using the zero-momentum
Ward identities (see eq.(2.14)) we find
$$
\DD \hat\C_{\phi_5^a NN}\big|_{k=0} = 
-\c_\phi \hat\C_{\phi_5^a NN}\big|_{k=0}
+ \c \Phi_{a0}^{-1} \Phi_{0b} \hat\C_{\phi_5^b NN}\big|_{k=0}
\eqno(C.12)
$$
Since $\hat\C_{\phi_5^a NN}\big|_{k=0} = 
B_{ab}^T \hat \C_{\eta^b NN}\big|_{k=0}$, and comparing eq.(C.12)
with the RGE (C.8) for $B$, we confirm
$$
\DD \hat\C_{\eta^a NN}\big|_{k=0} = 0 ~~~~~~~~{\rm for~ all}~a
\eqno(C.13)
$$

\vskip1cm

\noindent{\bf Appendix D:~~ Comparison with the literature} 
\vskip0.5cm

In a number of papers and lectures (see e.g.~refs.[16,17]), Ioffe has
criticised our earlier work on the `proton spin', expressing his view
that our formal theory is ``not justifiable'' and claiming that the spectral
sum rule technique we use is not valid for calculating the topological
susceptibility or decay constants in the flavour singlet channel.

These criticisms are made explicit in the recent paper [17]. Describing
our derivation of the $U_A(1)$ Goldberger-Treiman relation (in the
chiral limit) for $G_A^0$, Ioffe states that ``the matrix element
$\langle p|Q|p\rangle$ was saturated by contribution of two operators
$Q$ and singlet pseudoscalar operator $\phi_5$ -- and the result was
obtained by orthogonalisation of the corresponding matrix.''
This simply reflects a failure to understand our theoretical method. 
As we have repeatedly emphasised, no
approximation is involved in the decomposition of the matrix element
into composite propagators and 1PI vertices, as e.g.~in eq.(3.5) here.
If a different basis of operators is chosen, the definition of the
vertices changes too -- they become 1PI with respect to the new basis.
The basis of operators chosen for the decomposition is in no sense required
to be `complete' -- the procedure is {\it not} the familiar quantum mechanical
one of inserting a complete set of states. The matrix element is {\it not} 
being saturated with a restricted number of operators chosen from some
complete set. The only approximation comes in our subsequent conjecture
that the vertices, defined specifically as we have defined them, obey
the OZI rule. The motivations and {\it a posteriori} justifications for this
conjecture are explained carefully and at length in our papers.

Still referring to our work[12], Ioffe continues, ``the calculation of
$\chi'(0)$ by QCD sum rules is not correct, because as shown in ref.[19]
by considering the same problem with account of higher order terms of OPE 
than it was done in [12], the OPE breaks down at the scales characteristic
of this problem.'' This refers to ref.[19], where Ioffe and Khodzhamiryan
suggested that the extent of $SU(3)$ breaking which occurs in the Laplace
sum rule approach is unrealistically large, and concluded that QCD spectral
sum rules were unreliable in the $U_A(1)$ channel.
In the rest of this appendix, we show explicitly how their calculation 
is related to ours and point out a number of problems (errors in the QCD 
expressions, inconsistencies of the input and stability parameters, etc.)
in their approach which are responsible for this false conclusion.

We consider the current correlation function $$\eqalignno{
\Pi_{\m\n}^{0q}(k) ~&=~ i \int d^4x ~e^{ik.x}~\langle0|T~J_{\m5}^0(x)~
J_{\n5}^q(0)|0\rangle \cr
&=~ \Pi_T^{0q}(k^2) (g_{\m\n} - {k_\m k_\n\over k^2}) + \Pi_L^{0q}(k^2)
{k_\m k_\n\over k^2} &(D.1) \cr } $$
where $J_{\m5}^{0q} = \bar q\c_\m\c_5 q$ is the axial current for each
flavour of quark $q=u,d,s$ separately. The notation follows our eq.(5.2).
Comparing with the notation of ref.[19], the $\Pi_L$ are the same up to a
minus sign, whereas our $\Pi_T$ is a linear combination of the form factors
defined there. Only $\Pi_L$ plays a role in what follows.

Taking the divergences, and using the chiral Ward identity (2.9), we have $$
k^\m k^\n \Pi_{\m\n}^{0q}(k) ~=~ k^2 \Pi_L^{0q}(k^2) ~=~ i\int
d^4x~e^{ik.x}~\langle0|T~D^0(x)~D^q(0)|0\rangle + 4 m_q \langle \bar q
q\rangle
\eqno(D.2)
$$
where
$$\eqalignno{
D^0 &= 2n_f Q + \sum_{q=u,d,s}~2m_q \bar q\c_5 q \cr D^q &= 2Q + 2m_q \bar
q\c_5 q &(D.3) \cr } $$
The relation with the correlators studied in the text is therefore $$
\psi_5(k^2) - \psi_5(0) ~=~
{1\over (2n_f)^2} k^2\sum_{q=u,d,s}~\Pi_L^{0q}(k^2) \eqno(D.4)
$$
with the other $SU(3)$ combinations giving $\psi_5^{03}$ and $\psi_5^{08}$.
Taking the derivative wrt $k^2$ and evaluating at $k=0$, we find $$
\psi'_5(0) ~=~ {1\over (2n_f)^2} \sum_{q=u,d,s}~\Pi_L^{0q}(0) \eqno(D.5)
$$
$\Pi_L^{0q}(k^2)$ obeys an unsubtracted dispersion relation $$
\Pi_L^{0q}(k^2) ~=~ \int_0^\infty {dt\over t-k^2-i\e} {1\over\pi} {\rm
Im}\Pi_L^{0q}(t)
\eqno(D.6)
$$
The corresponding Laplace sum rule is simply $$
\t^{-3} {\cal L}({\cal F}_\Pi) ~=~ \int_0^\infty dt e^{-t\t} {1\over\pi}
{\rm Im} \Pi_L^{0q}(t)
\eqno(D.7)
$$
We can also write a once-subtracted sum rule, which enables us to calculate
$\Pi_L^{0q}(0)$:
$$
\t^{-2} {\cal L}({\cal G}_\Pi) + \Pi_L^{0q}(0) ~=~ \int_0^\infty {dt\over
t} e^{-t\t} {1\over\pi} {\rm Im} \Pi_L^{0q}(t) \eqno(D.8)
$$
Here,
$$
{\cal F}_\Pi \equiv {d^2\over{(dK^2)^2}}\Pi_L^{0q} ~~~~~~~
{\cal G}_\Pi \equiv {d\over{dK^2}}\biggl({\Pi_L^{0q}\over K^2}\biggr)
\eqno(D.9)
$$
Taking the $SU(3)$ singlet combination of these sum rules for $q = u,d,s$
clearly gives just the Laplace sum rules (6.18) and (6.19) analysed in the
text.

In ref.[19], Ioffe and Khodzhamiryan define the `decay constants' for
individual quark flavours as follows: $$
\langle0|J_{\mu5}^q|\eta'\rangle = ik_\m g_{\eta'}^q \eqno(D.10)
$$
and by saturating the r.h.s. of eq.(D.7) with the $\eta'$, obtain sum rules
for $f_{\eta'}g_{\eta'}^q$ for $q=u,d$ and $q=s$. Taking the ratio, they
find an unrealistic $SU(3)$ breaking characterised by $g_{\eta'}^s /
g_{\eta'}^{u,d} \sim 2.5$ and conclude that the Laplace sum rule for
$\Pi_L$ with the flavour singlet axial current does not work.

However, as we have shown in the text, a correct implementation of the sum
rule method does indeed work, showing good stability over a range of
appropriate $\t$ and $t_c$ values and giving results for the decay
constants in both flavour singlet and non-singlet channels which show only
the expected level of $SU(3)$ breaking. In the rest of this
section, we shall therefore quote the full formula for the sum rule (D.7)
for $\Pi_L^{0q}$, correcting some mistakes and omissions in ref.[19], then
briefly indicate some of the problems with their calculation.

From the results given in section 6, we can immediately read off the
required sum rule:
$$\eqalignno{
2n_f\sqrt2 f_{\eta'} g_{\eta'}^q m_{\eta'}^2 e^{-\t m_{\eta'}^2}~+~
\ldots ~&=~ 
{3\over8\pi^2}\t^{-2} \aspb^2
\bigl(1-(1+t_c\t)e^{-t_c\t}\bigr)(1+\d_{gg}^{PT}) \cr &~~~~+ {3\over2}
\aspb \biggl[{1\over2\pi}\langle \a_s G^2\rangle (1 + \d_{gg}^{NP}) +
\aspb \t \langle gG^3\rangle \biggr]\cr
&~~~~+\d_{qs}{3\over2\pi^2} \t^{-1} m_s^2 (1- e^{-t_c\t}) (1+\d_{qq}^{PT})
\cr &~~~~- 2(1 + 3\d_{qs}) \aspb M_0^2\t m_s \langle \bar s s\rangle -
\d_{qs} 4 m_s \langle \bar s s\rangle \cr & \hfil &(D.11) \cr}$$
where (see also [35])
$$\eqalignno{
\d_{gg}^{PT}&=\aspb \biggl[{83\over 4} +\b_1(1-\c_E) + 6\Bigl(3m_s^2 +
{\pi\over{\bar{\alpha_s}}}\l^2\Bigr)\t\biggr] \cr
\d_{gg}^{NP}&=-\aspb\Bigl({9\pi\over 8}\Bigr)(1-e^{-t_c\t}) \cr
\d_{qq}^{PT} &= \Bigl({17\over3} + 2\c_E\Bigr) \aspb &(D.12) \cr} $$
The dots on the l.h.s. denote the presence of any further intermediate
states (in particular the $\eta$) which might be required. The terms
appearing with $\d_{qs}$ are present only when $q$ is chosen to be the $s$
quark, and as usual we have assumed $m_u = m_d = 0$.

\vskip0.2cm
We now comment on some of the differences between the paper [19] and our work:

\noindent $\bullet$ ~~It is {\it a priori} dangerous to use the sum rule
for $\Pi_L^{0q}$ with the individual quark flavours while only keeping the
$\eta'$ as an intermediate state in the sum rule. This would imply that
only the $\eta'$ and not the $\eta$ would contribute in the mixed $SU(3)$
combination $\Pi_L^{08}$.
It follows that the quantity $g_{\eta'}^s / g_{\eta'}^{u,d}$ used in
ref.[19] is not the most reliable measure of the strength of $SU(3)$
breaking: while a value close to 1 would indicate self-consistently that
$SU(3)$ symmetry is very accurate and the $\eta - \eta'$ mixing angle is
very small, a large value would simply show that the initial hypothesis of
$\eta'$ saturation for the individual $\Pi_L^{0q}$ is not self-consistent.
(Ref.[19] refers to a further calculation taking $\eta - \eta'$ mixing into
account, but the details are not given.)

\noindent $\bullet$ ~~The formula (D.11) differs in a number of important ways
from that used in ref.[19]. In (D.11), the coefficient of the term $4 m_s
\langle \bar s s\rangle$
coming from the zero-momentum Ward identities has been corrected compared
to ref.[19]. We have
included the leading-order radiative corrections, which in this example
have a significant effect in
inducing stability in $\t$ and $t_c$. Notice that the radiative corrections
have a far greater
effect on this single sum rule than in the combined sum rule used in the
text to calculate
$\psi'_5(0)$, where they cancel between the ${\cal L(F)}$ and ${\cal L(G)}$
terms. We have also
included the contribution of $O(m_s^2)$ arising from the correlator denoted
by $\psi_{qq}$ in the
text, which was omitted in ref.[19] but is of the same order as the other
principal terms for the values of $\t$ used in [19].

\noindent $\bullet$ ~~In our analysis of the sum rules, we find good stability
at values of $t_c \simeq 6 ~\GV^2$ and $\t$ in the range $(0.2 - 0.4)
~\GV^{-2}$. In contrast,
ref.[19] uses too low a value of $t_c\simeq$ 2.5 $\GV^2$ and without
finding stability takes an {\it ad hoc} $\t$ value of about
$ 1 ~\GV^{-2}$. This is far too high and goes against the expectation
[41,40] that the optimal
scale
$\t$ should be comparatively low in the gluon-rich flavour singlet
channels. For this value of $\t$,
the convergence of the OPE for the terms of $O(m_s^2\t)$ is also poor.

We can therefore conclude that the analysis of the sum rule for
$\Pi_L(k^2)$ in ref.[19] (equivalent to our once-subtracted sum rule for
$\psi_5(k^2)$) is
unreliable and that their conclusion that the spectral sum rule method does
not work in the flavour
singlet channel is mistaken.

\vskip0.3cm
By taking the once-subtracted sum rule (D.8) for $\Pi_L^{00}(k^2)$
in combination with (D.7), we can write a Laplace sum rule for
$\Pi_L^{00}(0)$, the quantity
required for the generalised Goldberger-Treiman relations (and therefore
the `proton spin' effect).
This sum rule is of course identical to that for $\psi'_5(0)$ analysed in
the text. This is not discussed in ref.[19]. 
Alternatively, Ioffe and Khodzhamiryan use a finite energy sum rule
(FESR), analogous to the one described in our eq.(6.45) but omitting the 
radiative corrections. However, the FESR should be used
with great care due to its strong $t_c$ dependence, and the fact that the
result comes from a difference of large numbers.
Indeed, it is unsurprising that by substituting their set of values of
$f_{\eta'}$, $g_{\eta'}$, and $t_c$, which are inconsistent with the values
we have obtained from the Laplace sum rule based on $\tau$ and $t_c$ stabilities, 
they obtain an unreliable result.

Finally, in more recent papers[20], Ioffe and Oganesian extend the 
work of ref.[19] by combining it with a sum rule for $G_A^0$ itself, derived 
using a composite 3-quark operator with the quantum numbers of the proton. 
However, one can immediately notice that the choice of the interpolating 
nucleon currents used by the authors is far away from the
optimised choice analysed in refs.[48,49].
Irrespective of the validity of this method, it is evident that the
results of ref.[20] are as unreliable as those discussed above, and for
the same reasons, since the results of ref.[19] are used as input parameters in 
the new sum rule.

\vskip1cm

\noindent{\bf Acknowledgements}
\vskip0.5cm
The work of one of us (G.M.S.) is partially supported by the EC TMR Network
Grant FMRX-CT96-0008 and by PPARC. He is grateful to the
Centre National de la Recherche Scientifique (CNRS) for hospitality
during his stay at the Universit\'e de Montpellier where part of this work
was done.

\vfill\eject

\noindent{\bf References}
\vskip0.5cm

\settabs\+\ [&1] &Author \cr

\+\ [&1] &G.M.~Shore, {\it Nucl.~Phys. (Proc.~Suppl.)} B 64 (1998) 167. \cr

\+\ [&2] &B.~Lampe and E.~Reya, hep-ph/9810270. \cr

\+\ [&3] &G.~Altarelli and G.G.~Ross, {\it Phys.~Lett.} B212 (1988) 391.  \cr

\+\ [&4] &R.D.~Ball, S.~Forte and G.~Ridolfi, {\it Phys.~Lett.} B378 (1996)
255.  \cr

\+\ [&5] &J.~Ellis and R.L.~Jaffe, {\it Phys.~Rev.} D9 (1974) 1444.  \cr

\+\ [&6] &G.~Altarelli and G.~Ridolfi, {\it Nucl.~Phys. (Proc.~Suppl.)} 39B,C 
(1995) 106. \cr

\+\ [&7] &G.~Veneziano, {\it Mod.~Phys.~Lett.} A4 (1989) 1605. \cr 

\+\ [&8] &G.~Veneziano, {\it in} `From Symmetries to Strings: Forty Years of
Rochester Conferences', \cr
\+\ &{} &~~~~~~~~~~ed. A.~Das, World Scientific, 1990. \cr 

\+\ [&9] &G.M.~Shore, {\it Nucl.~Phys. (Proc.~Suppl.)} 39B,C (1995) 101. \cr 

\settabs\+ [1&1] &Authors \cr

\+ [1&0] &G.M.~Shore and G.~Veneziano, {\it Phys.~Lett.} B244 (1990) 75. \cr 

\+ [1&1] &G.M.~Shore and G.~Veneziano, {\it Nucl.~Phys.} B381 (1992) 23. \cr 

\+ [1&2] &S.~Narison, G.M.~Shore and G.~Veneziano, {\it Nucl.~Phys.}
B433 (1995) 209.\cr 

\+ [1&3] &G.M.~Shore and G.~Veneziano, {\it Nucl.~Phys.} B516 (1998) 333.  \cr

\+ [1&4] &D.~De Florian, G.M.~Shore and G.~Veneziano, Proc.~of 1997 Workshop  \cr
\+\ &{} &~~~~~~~~~on Physics with Polarised Protons at HERA, hep-ph/9711358. \cr

\+ [1&5] &G.M.~Shore, {\it Nucl.~Phys. (Proc.~Suppl.)} B 54A
(1997) 122. \cr 

\+ [1&6] &B.L.~Ioffe, {\it in} Proceedings, Ettore Majorana International
School of Nucleon Structure, \cr
\+\ &{} &~~~~~~~~~~Erice, 1995; hep-ph/9511401. \cr 

\+ [1&7] &B.L. Ioffe, Lecture at St.~Petersburg Winter School, hep-ph/9804328.\cr

\+ [1&8] &B.V.~Geshkenbein and B.L.~Ioffe, {\it Nucl.~Phys.} B166 (1980) 340. \cr 

\+ [1&9] &B.L.~Ioffe and A.Yu.~Khodzhamiryan, {\it Yad.~Fiz.} 55 (1992) 3045. \cr

\+ [2&0] &B.L.~Ioffe and A.G.~Ognesian, {\it Phys.~Rev.} D57 (1998) R6590; 
hep-ph/9810430  \cr

\+ [2&1] &B.~Zumino, {\it in} Lectures on elementary particles and quantum
field theory, \cr
\+\ &{} &~~~~~~~~~~Brandeis 1970, Vol.~2, 441, ed.~S.~Deser et al. (MIT
press, Cambridge MA).\cr

\+ [2&2] &A.V.~Efremov, J.~Soffer, N.A.~T\"ornqvist, {\it Phys.~Rev.}
D44 (1991) 1369.   \cr

\+ [2&3] &H.~Leutwyler, {\it Nucl.~Phys. (Proc.~Suppl.)} B64 (1998) 223; \cr 
\+\ &{} &R.~Kaiser and H.~Leutwyler, hep-ph/9806336. \cr

\+ [2&4] &T.~Feldmann and P.~Kroll, {\it Eur.~Phys.~J.} C5 (1998) 327; \cr 
\+\ &{} &T.~Feldmann, P.~Kroll and B.~Stech, {\it Phys.~Rev.} D58 (1998) 11406;\cr
\+\ &{} &T.~Feldmann, hep-ph/9807367. \cr  

\+ [2&5] &SMC collaboration, {\it Phys.~Lett.} B412 (1997) 414. \cr  

\+ [2&6] &G.~Altarelli, R.D.~Ball, S.~Forte and G.~Ridolfi, {\it Nucl.~Phys.}
B496 (1997) 337.  \cr

\+ [2&7] &SMC collaboration, {\it Phys.~Rev.} D58 (1998) 112001\cr 

\+ [2&8] &P.~Di Vecchia and G.~Veneziano, {\it Nucl.~Phys.} B171 (1980) 253.\cr

\+ [2&9] &S.~Narison, {\it QCD spectral sum rules},
Lecture Notes in Physics, Vol 26 (1989)  \cr
\+\ &{} &~~~~~~~~~~ed.~World Scientific; {\it Recent Progress in QCD 
spectral sum rules}, (to appear).  \cr

\+ [3&0] &E.~Braaten, S.~Narison and A.~Pich, {\it Nucl.~Phys.} B373
(1992) 581; \cr  
\+\ &{} &ALEPH collaboration, {\it Phys. Lett.} B307 (1993) 209.\cr

\+ [3&1] &S.~Bethke, {\it Nucl.~Phys.~(Proc. Suppl.)} B,C39 (1995);
B, A54 (1997).  \cr

\+ [3&2] &E.G.~Floratos, S.~Narison and E.~de Rafael, {\it Nucl. Phys.}
B155 (1979) 115.  \cr

\+ [3&3] &S.~Narison, {\it Phys.~Lett.} B358 (1995) 113.  \cr

\+ [3&4] &M.~Davier,  {\it in} Proceedings, TAU 96, Colorado (1996); \cr 
\+\ &{} &S.~Chen, {\it Nucl.~Phys. (Proc.~Suppl.)} B64 (1998) 265;  \cr
\+\ &{} &A.~Pich and J.~Prades, hep-ph/9804462;\cr
\+\ &{} &K.J.~Chetyrkin, J.H.~Kuhn and A.~Pivovarov, hep-ph/9805335. \cr

\+ [3&5] &M.A.~Shifman, A.I.~Vainshtein and V.I.~Zakharov, {\it Nucl.~Phys.} 
B147 (179) 385, 448.  \cr

\+ [3&6] &S.~Narison, {\it Nucl.~Phys.} B509 (1998) 312.\cr

\+ [3&7] &G.M.~Shore and G.~Veneziano, {\it Nucl.~Phys.} B381 (1992) 3. \cr 

\+ [3&8] &S.~Narison and E.~de Rafael, {\it Phys.~Lett.} B103 (1981) 57.  \cr

\+ [3&9] &A.L.~Kataev, N.V.~Krasnikov and A.A.~Pivovarov, {\it Nucl.~Phys.} 
B198 (1982) 508; \cr
\+\ &{} &~~~~~~~~~~erratum hep-ph/9612326 (1996).  \cr

\+ [4&0] &K.G.~Chetyrkin, S.~Narison and V.I.~Zakharov, hep-ph/9811275.  \cr

\+ [4&1] &V.A.~Novikov et al., {\it Nucl.~Phys.} B237 (1984) 525. \cr

\+ [4&2] &D.~Asner et al., {\it Phys.~Lett.} B296 (1992) 171. \cr

\+ [4&3] &S.~Narison, {\it Phys.~Lett.} B361 (1995) 121; B387 (1996) 162. \cr

\+ [4&4] &M.~D'Elia, A.~Di Giacomo and E.~Meggiolaro, 
{\it Nucl.~Phys. (Proc.~Suppl.)} 64 (1998) 316. \cr

\+ [4&5] &S.~Narison, N.~Pak and N.~Paver, {\it Phys. Lett.} B147
(1984) 162.  \cr

\+ [4&6] &K.G.~Chetyrkin, D.~Pirjol and K.~Schilcher, hep-ph/9612394 (1996).\cr

\+ [4&7] &M.~Jamin and M.~Munz, {\it Z.~Phys.} {\bf C66} (1995) 633.\cr

\+ [4&8] &B.L.~Ioffe, {\it Nucl.~Phys.} B188 (1981) 317; B191 (1981) 591. \cr

\+ [4&9] &Y.~Chung, H.G.~Dosch, M.~Kremer and D.~Schall, {\it Phys.~Lett.} 
B102 (1981) 175; \cr
\+\ &{} &~~~~~~~~~~{\it Nucl.~Phys.} B197 (1982) 57;  \cr
\+\ &H.G.~Dosch, M.~Jamin and S.~Narison, {\it Phys.~Lett.} 
B220 (1989) 251.  \cr

\+ [5&0] &S.~Narison, {\it Phys.~Lett.} B210 (1988) 238.  \cr

\+ [5&1] &A.~Hoecker, {\it in} Proceedings, TAU 96, Colorado.   \cr

\+ [5&2] &V.A.~Novikov, M.A.~Shifman, A.I.~Vainshtein and V.I.~Zakharov, \cr
\+\ &{} &~~~~~~~~~~{\it Nucl.~Phys.} B191 (1981) 301 \cr

\+ [5&3] &G.~Boyd, B.~All\'es, M.~D'Elia and A.~Di Giacomo, {\it in} Proceedings,
HEP 97 Jerusalem, \cr
\+\ &{} &~~~~~~~~~~hep-lat/9711025.  \cr

\+ [5&4] &D.~Espriu and R.~Tarrach, {\it Z.~Phys} C16 (1982) 77. \cr

\vfill\eject

\noindent{\bf Figures}

\vskip1cm

\centerline{
{\epsfxsize=10cm\epsfbox{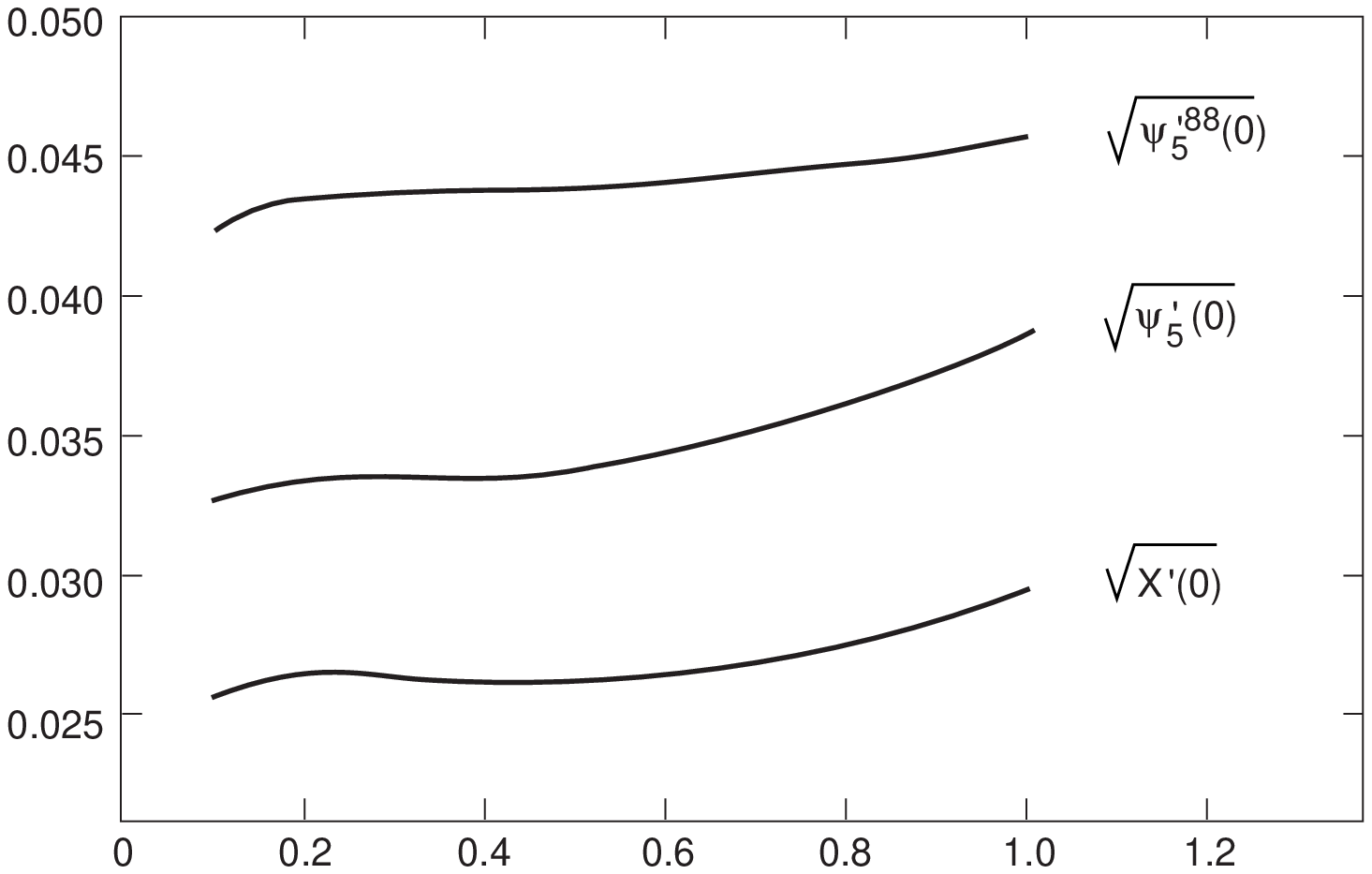}}}
\vskip0.2cm
\noindent{\bf Fig.~1}~~The dependence of the correlation functions
$\chi'(0)$, $\sqrt{\psi'_5(0)}$ and $\sqrt{\psi_5^{\prime 88}(0)}$ 
(in $\GV$) on the Laplace sum rule parameter $\t$ (in $\GV^{-2}$).

\vskip2cm

\centerline{
{\epsfxsize=10cm\epsfbox{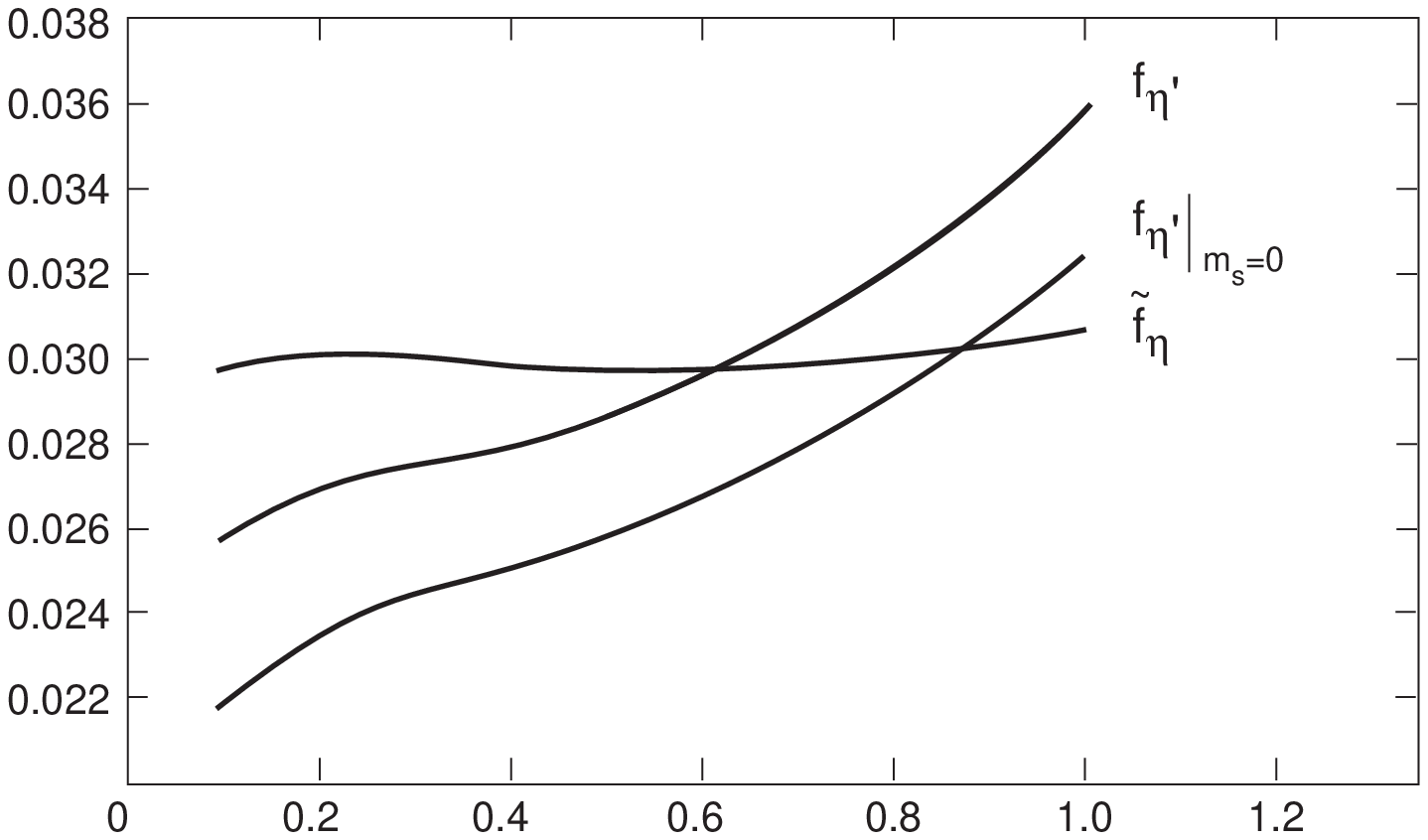}}}

\vskip0.2cm
\noindent{\bf Fig.~2}~~The dependence of the decay constants
$f_{\eta'}\big|_{m_s=0}$, $f_{\eta'}$ and $\tilde f_{\eta}$ (in $\GV$)
on the Laplace sum rule parameter $\t$ (in $\GV^{-2}$).

\bye